\documentclass[12pt]{article}

\usepackage[english]{babel}
\usepackage{rotating}
\usepackage{authblk}
\usepackage{url}
\usepackage[letterpaper,top=2.5cm,bottom=2.5cm,left=2.5cm,right=2.5cm,marginparwidth=1.75cm]{geometry}

\usepackage{amsmath}
\usepackage{ amssymb }
\usepackage{graphicx}
\usepackage{xcolor, colortbl}
\usepackage{lscape}
\usepackage{array, multirow, multicol}
\usepackage[colorlinks=true, allcolors=blue]{hyperref}
\usepackage{fancyhdr}
\usepackage{cleveref}
\usepackage[backend=biber, sorting=none]{biblatex} 
\usepackage{subfigure}
\usepackage{mathtools}
\usepackage{cancel}
\usepackage{dsfont}
\usepackage{orcidlink}
\newcommand{\eq}[1]{\begin{align}#1\end{align}}

\renewcommand{\u}[1]{\textrm{U}(#1)}
\newcommand{\su}[1]{\textrm{SU}(#1)}

\newcommand{\nn}{\nonumber} 

\newcommand{\ba}{\begin{array}}
\newcommand{\ea}{\end{array}}

\newcommand{\lag}{{\cal L}}

\newcommand{\hc}{{\rm h.c.}}
\hypersetup{colorlinks, citecolor=blue, linkcolor=newred, urlcolor=blue}
\definecolor{newred}{rgb}{0.68, 0.15, 0.20}
\definecolor{newgray}{rgb}{.85,.85, .85}

\DeclareUnicodeCharacter{2212}{-}

\title{Dark Radiation production in axion misalignment mechanisms and the cosmological tensions}

\author[1,2]{José María Pérez-Poyatos} 
\author[2]{Ver\'onica Sanz}
\affil[1]{Universidad de C\'ordoba, E-14071 C\'ordoba, Spain}
\affil[2]{Instituto de F\'isica Corpuscular (IFIC), Universidad de Valencia-CSIC, E-46980 Valencia, Spain}
\date{}
\begin{document}
\maketitle

\begin{abstract}
    We investigate the production of dark radiation (DR) from axions and axion-like particles (ALPs) as potential origins of dark matter. Focusing on the dark matter misalignment mechanism, we examine non-thermal, pre-inflationary scenarios that could lead to the generation of DR. A key part of our analysis involves a Bayesian approach to confront ALP parameter space with current cosmological data. Additionally, we explore whether DR production could offer solutions to persistent cosmological anomalies, particularly those related to the Hubble constant $H_0$ tension and the $S_8$ parameter discrepancy. Our findings aim to shed new light on the role of ALPs in addressing these open issues in cosmology.

\end{abstract}

\section{Introduction}

The prospect of axions or axion-like particles  serving as the solution to the enigma of dark matter in the universe presents an intriguing avenue within particle physics. Furthermore, the emergence of axions from a phase transition in the early universe holds the potential to alter the vacuum structure and contribute to important phenomena such as the production of observable gravitational waves. 

The QCD axion was initially proposed to resolve the Strong CP problem, addressing the absence of CP violation in strong interactions. To tackle this issue, an extension to the Standard Model (SM) gauge group, $\su3_c\times\su2_L\times \u1_Y$, is introduced by incorporating a global $\u1$ group \cite{Peccei:1977hh,Wilczek:1977pj,Weinberg:1977ma,Dine:1981rt}. This extension undergoes spontaneous breaking facilitated by the vacuum expectation value of a complex scalar field,  singlet under the SM gauge symmetries,  at the energy scale $f_\phi$. Consequently, the axion arises as the (CP-odd) Goldstone boson linked to this breaking mechanism.

However, the global \u1 is broken at the quantum level, and the axion develops small (and temperature-dependent) mass and anomalous couplings to the SM gauge fields. Regarding fermions, the axion's couplings are derivative, thereby preserving the Goldstone nature of the axion. These couplings emerge at dimension 5 and are suppressed by the scale $f_\phi$.

The scale of spontaneous symmetry breaking $f_\phi$ was initially proposed to be around the electroweak scale $v\sim 246$~GeV. However, the lack of signals of the axion implies that it must be very weakly coupled to the SM particles which in turn sets $f_\phi>10^9$~GeV: the so called invisible axion paradigm (see current bounds and future detection prospects~\cite{cajohare,Eby:2024mhd}). 

On the other hand, models that try to address other theoretical problems including Composite or Little Higgs~\cite{DelAguila:2019xec,Illana:2021uwu,Illana:2022wqv,Croon:2015fza}, Extra Dimensions~\cite{Croon:2014dma}, Majorons ~\cite{Cuesta:2021kca,Cuesta:2023awo}, or more generally string theory~\cite{Svrcek:2006yi} also include in their spectrum scalar particles resulting from the spontaneous breaking of global symmetry groups that exhibit properties akin to those of the QCD axion: they are singlets under the SM gauge group, naturally lightweight, and weakly coupled. They are commonly referred to as \textit{axion-like particles} (ALPs). 

The intrinsic properties of axion-like particles (ALPs) render them promising candidates for addressing cosmological and astrophysical issues such as inflation (in the natural inflation scenario), protect the inflationary potential from dangerous quantum corrections \cite{Freese:1990rb,Adams:1992bn,Kim:2004rp,Dimopoulos:2005ac,Maleknejad:2011jw,Adshead:2012kp,Peloso:2015dsa}, explain particle production in the early universe 
\cite{Anber:2009ua,Cook:2011hg,Barnaby:2011qe,Croon:2015naa,Namba:2015gja,Dimastrogiovanni:2016fuu,Peloso:2016gqs,Garcia-Bellido:2016dkw,Domcke:2016bkh}, the origin of pulsar timing array measurements \cite{Unal:2023srk} and the dark matter (DM) problem, see e.g.~\cite{Marsh:2015xka}. 

Various mechanisms exist for producing DM from ALPs. In this study, we confine our focus to non-thermal dark matter production mechanisms within the pre-inflationary framework, where the global symmetry responsible for ALP emergence is broken before inflation and remains so. Non-thermal mechanisms present an appealing alternative to the conventional thermal Weakly Interacting Massive Particle (WIMP) paradigm. A notable non-thermal mechanism is the so called realignment or conventional misalignment mechanism~\cite{Preskill:1982cy, Dine:1982ah,Abbott:1982af,Kolb:1990vq}. In the early universe, ALPs may deviate from the CP-even minimum of their potential. When the Hubble parameter $H$ becomes comparable to the ALP mass $m_\phi$ at the temperature $T_*$ determined by $m_\phi=3H(T_*)$, the ALP initiates oscillations around the minimum. These coherent oscillations act akin to matter and can yield the observed DM abundance. Pre-inflationary mechanisms circumvent issues like cosmic strings and domain walls, although they are subject to isocurvature constraints~\cite{Axenides:1983hj,Steinhardt:1983ia,Steinhardt:1984jj,Linde:1985yf,Seckel:1985tj,LYTH1990408,Turner:1990uz,Sikivie:2006ir,Beltran:2006sq,Kawasaki:2008sn,Kawasaki:2014una,Chen:2023txq}. 

Other mechanism that recently got more attention is the so called \textit{kinetic misalignment} (KM) mechanism \cite{Co:2019jts}. In this scenario the ALP has a non-zero initial velocity in field space before inflation. The expansion of the universe acts as a friction term, redshifting its velocity. If the ALP kinetic energy still dominates over the potential energy during radiation domination era at the temperature $T_*$, the ALP can explore several potential minima until kinetic energy redshifts to the height of the potential barrier. At that time, if the mass is larger than the Hubble expansion rate, the ALP starts oscillating around the minimum. As in the conventional misalignment mechanism
, the oscillations are responsible of generating DM.

On the other hand, Cosmology is currently living a high precision era. As a consequence some discrepancies are emerging between different datasets. Probably one of the best known is the $H_0$ tension, an $\sim 5\sigma$ discrepancy between inferences of the today Hubble expansion rate using early and late time universe datasets. This is largely driven by the discrepancy between the local measurement of the SH0ES collaboration using supernovas \cite{Verde:2019ivm,Knox:2019rjx,DiValentino:2021izs,Shah:2021onj} and
the model-dependent inference from the CMB and/or large-scale structure data using the $\Lambda\rm CDM$ model. A less stringent discrepancy is the $S_8$ tension, a discrepancy of $\sim 2-3 \sigma$ in the amount of clustering of matter
seen between late-universe datasets and the CMB
data \cite{Troster:2019ean,Ivanov:2019pdj,KiDS:2020suj,Hildebrandt:2020rno,KiDS:2020ghu,DES:2021wwk,Philcox:2021kcw,Krolewski:2021yqy,Garcia-Garcia:2024gzy,Nunes:2021ipq}. 

The purpose of this work is twofold. By considering that the ALP has a coupling to a dark radiation sector (DR) during matter domination epoch, we can first constrain the ALP parameter space confronting our model to recent cosmological data, and second, explore the possibility of solving the persisting $H_0$ and $S_8$ tensions.

The paper is structured as follows: Section~\ref{review} we review the different mechanisms for misalignment proposed in the literature. In Section~\ref{DRproduction} we explain our model in detail and provide analytical expressions for the production of dark radiation generated by the oscillations of the ALP around one of the minima of its potential. In section~\ref{Bayesiananalysis} we perform a Bayesian analysis to confront the ALP decaying model to current cosmological data to constrain the ALP parameter space and explore if it can indeed solve the $H_0$ and $S_8$ tensions. Finally, the concluding section~\ref{conclusions} summarizes our findings and presents our conclusions.

\section{Review on misalignment mechanisms  and cosmological tensions}\label{review}

In this section, we review the various misalignment mechanisms proposed for axion-like particles (ALPs) as a candidate for dark matter. These mechanisms are crucial for explaining how ALPs can acquire the correct relic density in the pre-inflationary scenario. We also discuss the cosmological tensions that arise, particularly in relation to the Hubble constant and the matter power spectrum. ALPs may offer potential resolutions or deepen these tensions depending on their interactions with other fields and their production history.

\subsection{Misalignment mechanisms}

In this section we briefly review some of the different misalignment mechanisms that have been proposed in the literature to produce ALP DM in the pre-inflationary scenario. These mechanisms are highly non-thermal. After the spontaneous breaking of the global symmetry that gives rise to the ALP, this chooses (probably by some stochastic process) a initial value $\theta_i = \left[-\pi, \pi\right.)$ in each Hubble patch. Later on, inflation homogenizes the value of the ALP in the whole universe. Typically when the ALP mass is of the order of the Hubble scale, the ALP oscillates around one of the minima of its potential. These oscillations are dumped by the expansion of the universe, relaxing the initial ALP field value to zero. The oscillations behave as a classical condensate of ALPs with zero momentum and thus can be the DM of the universe as we will show later.

Some of the proposed misalignment mechanisms are the following:

\begin{itemize}
    \item[1.] Conventional misalignment mechanism \cite{Preskill:1982cy, Dine:1982ah,Abbott:1982af}. The ALP is essentially frozen at the initial value $\theta_i$ due to the large Hubble friction. During radiation domination epoch, when the ALP mass becomes comparable to the Hubble expansion rate at the time $t_{\rm osc}$, $m_\phi\approx 3H(t_*)$, the ALP starts oscillating around one of the minima of the potential. For an ultralight QCD axion, it has been argued that the initial misalignment angle required to reproduced the current DM relic density is too small $\theta_i\sim \mathcal{O}\left(10^{-3}-10^{-4}\right)$, leading to a fine-tuning problem. However, in refs.~\cite{Co:2018phi,Co:2018mho} the authors propose a mechanism that allows to choose initial values closed to zero or to $\pi$ without fine-tuning. For example, Higgs inflation dynamics can ensure a large initial axion mass and thus stochastically the axion is relaxed to the minimum of the potential even during inflation. Together with a CP phase shift due to an approximate CP symmetry and a sign flip, the axion can be set to the bottom or the hilltop of the potential, and thus opening the window of light and heavy axion DM. This also helps to relax the isocurvature constraints afflicting the pre-inflationary axion.  
    
    \item[2.] Kinetic misalignment \cite{Co:2019jts,Barman:2021rdr}. In this case, the ALP has a non-vanishing initial velocity in field space, $\dot{\phi}_0$. If the kinetic energy associated to this field velocity is larger than the  potential energy at the time $t_{\rm osc}$, the ALP simply overcomes the potential barrier until the kinetic energy redshifts and the ALP gets trapped by the potential. As a consequence, the oscillations are delayed with respect to the previous scenario. If the initial ALP velocity is not sufficient to jump over a single potential barrier, this mechanism reduces to the conventional misalignment mechanism.
    \item[3.] Trapped scenario \cite{Nakagawa:2020zjr,DiLuzio:2021gos,Kitajima:2023pby}. In this scenario, two different potential minima develop during the universe evolution from high to low temperatures. Oscillations initially start around a false vacuum $\phi=\phi_*$, due to the temperature dependent potential, allowing $m_{\phi_*}(T)=d^2V(\phi)/d\phi^2 \gtrsim H(T)$ at $\phi_*$. Later on, the true minimum ($\phi=\phi_{\rm min}$) develops, the ALP mass becomes the vacuum mass $m_{\phi_{\rm min}}$, and the false vacuum becomes a maximum. Oscillations around the true minimum start whenever the ALP kinetic energy becomes smaller than the height of the potential barrier and the condition $m_{\phi_{\rm min}}>3H$ is fulfilled. These two stages are thus separated by a strongly non-adiabatic modification of the ALP potential.
    \item[4.] Frictional misalignment \cite{Papageorgiou:2022prc}.  In this case there is an sphaleron-induced thermal friction that effectively contributes to the Hubble friction $H$. Thermal friction can either enhance the axion relic density by delaying the onset of oscillations or suppress it by damping them.
\end{itemize}
Compared to conventional misalignment, in the rest of misalignment mechanisms the onset of the oscillations is delayed. However, once the oscillations have begun, the ALP behaves as matter in all the scenarios. This allows us to study their cosmological implications in a model independent fashion.

\subsection{Cosmological tensions}
In this section we briefly review the cosmological tensions which will play a relevant role in our discussion of the ALP parameters, the Hubble tension and the $S_8$ tension.

\subsubsection{The $H_0$ tension}

Using CMB data, the $\Lambda$CDM model predicts $H_0 = 67.4 \pm 0.5$~km/s/Mpc \cite{Planck:2018vyg}. This result is obtained by directly measuring the angular size of the acoustic scale in the CMB power spectrum. In contrast, local measurements of $H_0$, determined through a distance-redshift relationship (known as the ``cosmic distance ladder”), are in tension with this prediction. For instance, the latest measurement from the SH0ES collaboration reports $H_0 = 73.04 \pm 1.04$~km/s/Mpc \cite{Riess:2021jrx}.

The CMB was released at recombination and serves as a probe of the early universe. The estimation of $H_0$ from CMB data relies on the precise measurement of the acoustic scale imprinted on the CMB by baryon acoustic oscillations (BAOs), our model of the universe's expansion history, aka $\Lambda$CDM, and the model of the sound horizon at recombination. The acoustic scale in the CMB, $\theta_s$, represents the projection of the physical sound horizon at recombination, $r_s(z^*)$, which is the distance a sound wave could travel from the beginning of the universe until recombination according to

\eq{\label{sound_horizon}
\theta_s=\frac{r_s(z^*)}{D_A(z^*)}
}
where $D_A(z^*)$ is the comoving angular diameter distance to the surface where the CMB was released at redshift $z^*$
(the “surface of last scattering”). The sound horizon at recombination is given by
\eq{
r_s(z^*)=\int_{z^*}^{\infty}\frac{dz}{H(z)}c_s(z),
}
where $c_s(z)$ is the sound speed, while the comoving distance is given by
\eq{
D_A(z^*) = \int_0^{z^*}\frac{dz}{H(z)}.
}
Using Einstein equations for a spatially flat FLRW universe, $H(z)$ is fully determined by the Friedmann equation
\eq{\label{friedmann}
H(z) = \sqrt{\frac{\rho(z)}{3 M^2_p}},
}
where $\rho(z)$ is the sum of the energy densities of all the species present in the universe at redshift $z$. In the $\Lambda\rm CDM$ model
\eq{\label{all_densities}
\rho(z)=\rho^0_m(1+z)^3 + \rho^0_\gamma (1+z)^4 + \rho_\nu(z) + \rho_\Lambda,
}
where the superscript 0 indicates the present, $m$ denotes the sum of baryons and DM, $\gamma$ denotes photons and massless neutrinos, $\nu$ refers to massive neutrinos and $\Lambda$ the cosmological constant. $\rho_m^0$ is constrained indirectly by the CMB that mostly constrains $\rho_m(z\approx z^*)$ and extrapolating until today using the standard evolution of matter, $\rho^0_\gamma$ is constrained by the monopole temperature of the CMB. $\rho_{\nu}$ is also constrained by the CMB even though its evolution depends on the neutrino masses. Only remains the energy density of dark energy $\rho_\Lambda$. However, the CMB provides a direct constraint on $\theta_s$ and the physics governing the sound speed $c_s(z)$ is well understood within $\Lambda\rm CDM$. Thus, the CMB data combined with eq.~(\ref{sound_horizon}) specify $\rho_\Lambda$. Using eqs.~(\ref{friedmann}) and (\ref{all_densities}), $H(z)$ and thus $H_0 \equiv H(z = 0)$ can be inferred from the $\Lambda\rm CDM$ model. 

On the other hand, the direct measurement of $H_0$  is model-independent and serves as a probe of the late-time universe. This measurement is based on the recession velocity and distance to distant objects. The velocity is obtained from the redshift spectra of objects, while the distance requires the use of standard candles with known intrinsic brightness, along with the luminosity-distance relation. Type Ia supernovae (SNe) serve as standard candles for measuring $H_0$, though they only constrain $ H(z)/H_0$, since their absolute brightness normalization is not known. To determine their intrinsic brightness, we need absolute distance measurements for some of the SNe, which can be obtained using Cepheid variables. Cepheid brightness is calibrated through parallax measurements of nearby Cepheids in our galaxy neighborhood. This method forms the ``cosmic distance ladder": parallax measurements of nearby Cepheids calibrate more distant Cepheids, which in turn calibrate nearby SNe, allowing for the calibration of farther SNe. $H_0$ is measured using this method.

This discrepancy between the predicted value of $H_0$ using the $\Lambda$CDM model and the measured value using supernovae, might indicate the presence of new physics between the era of recombination and the present day. Several solutions to the Hubble tension has been already proposed in the literature \cite{Schoneberg:2021qvd}. For instance, there are models that feature extra relativistic relics, with different assumptions concerning interactions in the sectors of dark radiation neutrinos and dark matter; models in which the sound horizon is shifted due to some other ingredient such a primordial magnetic field; or models with a modification of the
late-time cosmological evolution, either due to Dark Energy or late
decaying Dark Matter.

\subsubsection{The $S_8$ tension}
The $S_8$ parameter is defined as
\eq{
S_8\equiv\sigma_8\left(\frac{\Omega_m}{0.3}\right)^{\frac{1}{2}},
}
where $\Omega_m\equiv \rho^0_m/\rho^0_{\rm crit}$ is the today density of matter normalized to the critical density and $\sigma_8$ is the root mean square of matter fluctuations over a sphere of radius $R = 8 $~Mpc/h at $z = 0$:
\eq{
(\sigma_8)^2=\frac{1}{2\pi^2}\int\frac{dk}{k}W^2(k R) k^3 P(k),
}
where $P(k)$ is the linear matter power spectrum today and $W(kR)$ is a spherical top-hat filter of radius $R =8$~Mpc/h.

An increasing tension has recently emerged between the value of $S_8$ measured from late-universe observations and the value inferred indirectly from the CMB by first constraining the parameters $\Lambda\rm CDM$ and then calculating $S_8$. Specifically, weak lensing surveys like KIDS yield $S_8=0.759\pm 0.024S8$ \cite{KiDS:2020suj}, while clustering survey analyses, such as those from BOSS, similarly report lower $S_8$ values \cite{Philcox:2021kcw,Ivanov:2019pdj}. The DES survey measures $S_8=0.776\pm 0.017$ from galaxy-galaxy lensing \cite{DES:2021wwk}, that is, a combined analysis of foreground galaxy clustering and background galaxy lensing. These measurements contrast with the indirect Planck constraint of $S_8=0.834\pm 0.016$ derived from the primary CMB \cite{Planck:2018vyg}.

\section{Pre-inflationary ALP decaying to dark radiation}\label{DRproduction}

We will consider a model with a classical ALP field that, after oscillating around one of the minima of its potential during radiation domination epoch, behaves as usual cold dark matter (DM) $\rho_{\phi}\sim a^{-3}$. We call $a_{\rm osc}$ to the scale factor of the universe at the moment when the ALP begins to oscillate. The ALP will dominate the energy content of the universe around recombination. Later on, a coupling of the ALP to a dark radiation (DR) sector activates when the scale factor is $a_{\rm DR}$. This implies that the ALP energy density will dilute faster than usual DM since a part of it will become DR. Hence, this model presents a new variation of the decaying cold dark matter to dark radiation scenario ~\cite{Poulin:2016nat,Chudaykin:2016yfk,Chudaykin:2017ptd,Nygaard:2020sow}. 
An alternative modification of this scenario is presented in ref.~\cite{Bringmann:2018jpr}, where a phenomenological formula for the DM evolution is proposed. As a consequence, the decay width of the converting part of the DM into DR is time dependent. Its cosmological implications are examined in refs.~\cite{Bringmann:2018jpr,DES:2020mpv,McCarthy:2022gok}. The main difference with respect to our work is that our formula will be derived from first principles as we will show along this section. Additionally, there are models that consider that the DM is composed by axions with a large velocity dispersion (warm DM) decaying into DR~\cite{Buen-Abad:2019opj}, which also aim to address the $S_8$ and $H_0$ cosmological tensions.

In the following we present in detail the main equations and develop the formalism that we will use to obtain the energy density of ALPs and DR.

\subsection{Background cosmology}
We assume that the universe is well described by the spatially flat FLRW metric
\eq{
ds^2 = g_{\mu\nu}dx^\mu dx^\nu = dt^2-a^2(t)d\vec{x}^2,
}
in comoving coordinates. In most of this work we will use conformal time, defined as follows
\eq{
dt\equiv a(t)d\eta
}
such that the FLRW metric takes the form

\eq{
ds^2=a^2(\eta)\left(d\eta^2-d\Vec{x}^2\right).
}

In this work we assume the standard cosmological history of the universe with the inclusion of the homogeneous ALP field that will be our DM candidate, as we will justify in the next subsection. The physics previous to the release of the CMB are well understood and explained within the $\Lambda\rm CDM$ model. Hence, any modification to $\Lambda\rm CDM$ induced by the ALP dynamics must enter into the game after recombination, at redshift $z\approx 1100$, when the universe is approximately in matter domination era. Hence, the Hubble expansion rate $H$ as a function of time is given by
\eq{
H\equiv\frac{\dot{a}}{a}=\frac{2}{3t}
}
where dot derivatives are taken with respect to cosmic time, $a$ is the scale factor and the last equality is valid in a matter dominated universe, when we assume that dark radiation is produced. It is convenient to integrate this expression to obtain the scale factor as a function of time
\eq{\label{scalefactor}
a(t)=a_0\left(\frac{t}{t_0}\right)^{\frac{2}{3}}\equiv\left(\frac{t}{t_0}\right)^{\frac{2}{3}}
}
where the subscript $0$ means today and $a_0=1$. Since the universe has only recently entered the dark energy-dominated epoch, this is a good approximation for the purposes of this work. The Hubble expansion rate as a function of the scale factor is
\eq{\label{Hubble}
H=\frac{2}{3t_0 a^{\frac{3}{2}}}=H_0 a^{-\frac{3}{2}},
}
where $H_0$ is the Hubble expansion rate today.

\subsection{ALP equation of motion}

Let us study the behavior of the homogeneous ALP field $\phi(\eta)=f_{\phi}\theta(\eta)$ in the expanding universe. Its equation of motion (eom) comes from the action of an scalar field minimally coupled to gravity
\eq{
S_{\phi}=\int d^4x \sqrt{\lvert g \rvert}\left(\frac{1}{2}\phi'^2-V(\phi)\right),
}
where $\lvert g\rvert = a^4(\eta)$ is the determinant of the metric $g_{\mu\nu}$, prime derivatives are taken with respect to conformal time and $V(\phi)$ is defined in analogy to the QCD axion
\eq{
V(\phi)\equiv m_\phi^2 f_\phi^2\left[1-\cos\left(\frac{\phi}{f_\phi}\right)\right],
}
but for simplicity we choose the ALP mass $m_\phi$ to be constant. Defining the dimensionless field $\theta \equiv \phi/f_\phi$, the ALP eom is thus given by
\eq{\label{alpeom}
\theta'' + 2\frac{ a'}{a}\theta' + a^2 m_\phi^2\sin\theta=0\rightarrow \theta'' + 2H a\theta' + a^2 m_\phi^2\sin\theta=0.
}
At late times but during radiation domination era, when $m_\phi \gtrsim 3H$,  the ALP starts oscillating with respect to one of the minima of the potential, namely $\theta_{\rm min}$. Thus the ALP can be redefined as $\theta=\theta_{\rm min}+\delta\theta$ where $\delta\theta$ is the displacement with respect to the oscillation minimum. The solution of the eom when $\delta\theta\lesssim 1$ such that $\sin\theta\approx \delta\theta$, is
\eq{
\label{alpposition}&\delta\theta=\delta\theta_{\rm i} \left(\frac{a_{\rm i}}{a(\eta)}
\right)^{\frac{3}{2}}\sin\left[m_\phi t(\eta)+\nu\right],\\
\label{alpvelocity}&\theta'=\delta\theta_{\rm i} a^{\frac{3}{2}}_{\rm i}a^{-\frac{1}{2}}(\eta)m_\phi\left[-\frac{3}{2}\frac{H}{m_\phi}\sin(m_\phi t(\eta)+\nu)+\cos(m_\phi t(\eta)+\nu)\right]
}
where $\delta\theta_i$ and $\nu$ are fixed from the numerical solution of eq.~(\ref{alpeom}) and $a_{\rm i}=a(\eta_{\rm i})$ at some reference time when oscillations have already begun.  The energy density of the ALP field during oscillations, when we can safely neglect $H/m_\phi$, is
\eq{\label{ALPenergydensity}
\rho_\phi(a)=\frac{1}{2}f_\phi^2\dot{\theta}^2+\frac{1}{2}m_\phi^2f_\phi^2\delta\theta^2 = \frac{1}{2}m_\phi^2f_\phi^2\delta\theta_{\rm i}^2\left(\frac{a_{\rm i}}{a}\right)^3,
}
where se used $\dot{\theta}=\theta'/a$. Notice that after the onset of the oscillations the ALP energy density scales as $\rho_\phi\sim a^{-3}$, independently of the background cosmology since we neglected $H/m_{\phi}$. Analogously, the pressure is given by
\eq{
p_\phi(a)= \frac{1}{2}f_\phi^2\dot{\theta}^2-\frac{1}{2}m_\phi^2f_\phi^2\delta\theta^2 =\frac{1}{2}m_\phi^2f_\phi^2\delta\theta_{\rm i}^2\left(\frac{a_{\rm i}}{a}\right)^3\cos\left[2(m_\phi t + \nu)\right].
}
In a period $T=\pi/m_\phi$ the scale factor changes an amount
\eq{
\Delta a\approx \dot{a}\Delta t = H a \frac{\pi}{m_\phi}\rightarrow \frac{\Delta a}{a}=\pi\frac{H}{m_\phi},
}
that is negligible after the onset of the oscillations. Hence, averaging over a period we have $\langle \cos\left[2(m_\phi t + \nu)\right] \rangle_T = 0$ such that the pressure is
$p_\phi(a) = 0$. Hence the ALP behaves as cold matter when oscillations have begun, justifying that the ALP can be the DM that will dominate the energy content of the universe after recombination. Sufficient DM will be produced if oscillations begin well inside radiation domination epoch.

\subsection{Dark radiation production by ALPs}

\subsubsection{Occupation number}
In this work we will assume that the ALP possesses a trilinear coupling to a dark radiation sector, namely
\eq{
\lag_{\phi\rm DR}=-\frac{1}{8}g_{\phi\rm DR}\phi\frac{\epsilon^{\mu\nu\rho\lambda}}{\sqrt{\lvert g \rvert}}F_{\mu\nu}F_{\rho\lambda},
}
where $\epsilon^{\mu\nu\rho\lambda}$ is the totally antisymmetric Levi-Civita tensor with $\epsilon^{0123}=+1$ and $g_{\phi\rm DR}\sim f_\phi^{-1}$ is a dimensionful coupling. Perturbativity imposes $g_{\phi\rm DR}f_\phi<1$.   
Since we only want to change the physics from recombination onwards, this coupling is only activated from $z\gtrsim 1100$. This can be justified, for instance, by assuming that the coupling of ALPs to dark radiation is an effective coupling mediated by new heavy states which modulate an energy-dependent coupling. Examples of an ALP coupling to gauge bosons mediated by new heavy states abound in the literature, particularly by adding several families of heavy fields. See e.g., Ref.~\cite{Choi:2020rgn} for a recent review  on axion models.

The action of dark photons coupled to the ALP is
\eq{
S_{\rm DR}=\int d^4x \sqrt{\lvert g\rvert}\left(-\frac{1}{4}F_{\mu\nu}F^{\mu\nu}+\lag_{\phi\rm DR}\right),
}
where $F_{\mu\nu}$ is the dark photon strength tensor $F_{\mu\nu}=\partial_\mu A_\nu-\partial_\nu A_\mu$. Hence, from the point of view of the DR, the ALP is just a background field that continuously supports particle production. At leading order, we consider that the ALP simply follows the eom (\ref{alpeom}), and thus neglecting the backreaction of the dark photons on the ALP field.
This effect will be later included when we impose energy conservation.

In temporal gauge $A_0=0$, and integrating by parts this reduces to
\eq{\label{photonaction}
S_{\rm DR}\equiv \int d^4x~ \lag_{\rm DR}=\int d^4x \left(\frac{1}{2}A_i'^2-\frac{1}{2}\left[\left(\partial_i A_j\right)^2-\left(\partial_i A_i\right)^2\right]+\frac{1}{2}g_{\phi\rm DR}\phi'\epsilon^{ijk}A_i\partial_jA_k\right)
}
where $i,j,k=1-3$ are spatial indices. In order to quantize the theory, we write $A_i$ in terms of creation and annihilation operators
\eq{\label{quantizedphoton}
A_i=\int \frac{d^3k}{(2\pi)^3}\sum_{\lambda=\pm}\left(\epsilon_i(\Vec{k},\lambda)\mu_{k\lambda}(\eta)a_{\Vec{k}\lambda}e^{i\Vec{k}\Vec{x}}+\epsilon^{*}_i(\Vec{k},\lambda)\mu^{*}_{k\lambda}(\eta)a^{\dagger}_{\Vec{k}\lambda}e^{-i\Vec{k}\Vec{x}}\right),
}
where the mode functions $\mu_{k\lambda}$ only depend on the modulus of $\lvert\vec{k}\rvert=k$ due to the isotropy of the FLRW metric and $\lambda=\pm$ are the two polarizations of the photon. Creation and annihilation operator satisfy canonical commutation relations
\eq{\label{creationandannihilationoperators}
[a_{\Vec{k}\lambda},a^{\dagger}_{\Vec{k'}\lambda'}]=(2\pi)^3\delta^3(\Vec{k}-\Vec{k}')\delta_{\lambda\lambda'}.
}
Notice that the delta function is related to the space volume through
\eq{
\lim_{\vec{k}\rightarrow \vec{k}'}\delta^3(\vec{k}-\vec{k}')=\lim_{\vec{k}\rightarrow \vec{k}'}\int d^3x~e^{i(\vec{k}-\vec{k}')\vec{x}}= V(\rightarrow \infty),
}
and thus having energy units of $E^{-3}$. This implies that the operators creation and annihilation operators $a^{\dagger}_{k,\lambda}$, $a_{k\lambda}$ have dimension $E^{-\frac{3}{2}}$. Consequently, since $A_{i}$ has dimension of $E$, the mode functions have dimension $E^{-\frac{1}{2}}$. 

On the other hand, the circular polarization vectors obey
\eq{\label{polarizationvectorrelation}
\epsilon(\vec{k},\lambda)\cdot\epsilon^{*}(\vec{k},\lambda')=\delta_{\lambda \lambda'},\quad \vec{k}\cdot \epsilon(\vec{k},\lambda)=0,\quad \vec{k}\times \epsilon(\vec{k},\lambda)=-i k \lambda \epsilon(\vec{k},\lambda), \quad \epsilon(-\vec{k},\lambda)=\epsilon^*(\vec{k},\lambda)
}
where the last two relations come from the fact that the two linear polarization vectors and the direction of propagation $\vec{k}$ constitute a direct trihedral.

The classical eom of the photon field is thus given by
\eq{
A_i''-\partial_j^2 A_i+\delta_{ij}\partial_j^2 A_i-g_{\phi\rm DR}f_\phi\theta'\epsilon^{ijk}\partial_j A_k=0.
}
Using the relations in eq.~(\ref{polarizationvectorrelation}) the mode functions satisfy
\eq{\label{modefunctioneom}
\mu_{k\lambda}''+(k^2-\lambda g_{\phi\rm DR}f_\phi k\theta')\mu_{k\lambda}=0,
}
which is the eom of a harmonic oscillator with a time-dependent frequency given by
\eq{\label{omega}
\omega^2_{k\lambda}=k^2-k\lambda g_{\phi\rm DR}f_\phi \theta'.
}
In the following we will derive physical arguments to fix the two initial conditions for the mode functions.

The first initial condition comes from imposing canonical commutation relations between the dark photon field $A_i$ and its corresponding conjugated momentum $\Pi_i$. This can be obtained trivially from the action in eq.~(\ref{photonaction})
\eq{\label{conjugatedmomentum}
\Pi_i=\frac{\partial\lag_\gamma}{\partial A_i'}=A_i'=\int \frac{d^3k}{(2\pi)^3}\sum_{\lambda=\pm}\left(\epsilon_i(\Vec{k},\lambda)\mu'_{k\lambda}(\eta)a_{\Vec{k}\lambda}e^{i\Vec{k}\Vec{x}}+\epsilon^{*}_i(\Vec{k},\lambda)\mu'^{*}_{k\lambda}(\eta)a^{\dagger}_{\Vec{k}\lambda}e^{-i\Vec{k}\Vec{x}}\right).
}
The canonical commutation relation at equal times imposes
\eq{\label{canonicalcommutator}
\left[A_i(\eta,\vec{x}),\Pi_j(\eta,\vec{y})\right]\equiv i\delta_{ij}\delta^3_{\rm tr}(\vec{x}-\vec{y}),
}
where the \textit{transverse} delta is defined through
\eq{
\delta_{ij}\delta^3_{\rm tr}(\vec{x}-\vec{y})\equiv\int\frac{d^3k}{(2\pi)^3}e^{i\vec{k}(\vec{x}-\vec{y})}\left(\delta_{ij}-\frac{k_ik_j}{k^2}\right).
}
Using the expansion of the dark photon field in eq.~(\ref{quantizedphoton}) and the relations in eqs.~(\ref{creationandannihilationoperators}) and (\ref{polarizationvectorrelation}), we arrive to
\eq{
\left[A_i(\eta,\vec{x}),\Pi_j(\eta,\vec{y})\right]=\int\frac{d^3k}{(2\pi)^3}\sum_{\lambda=\pm}\left(\epsilon_i(\vec{k},\lambda)\epsilon^*_j(\vec{k},\lambda)\mu_{k\lambda}\mu'^*_{k\lambda}e^{i\vec{k}(\vec{x})-\vec{y}}-\epsilon_i(-\vec{k},\lambda)\epsilon^*_j(-\vec{k},\lambda)\mu^*_{k\lambda}\mu'_{k\lambda}e^{-i\vec{k}(\vec{x}-\vec{y})}\right).
}
For the second term we change $\vec{k}\rightarrow-\vec{k}$, such that
\eq{
\left[A_i(\eta,\vec{x}),\Pi_j(\eta,\vec{y})\right]=\int\frac{d^3k}{(2\pi)^3}e^{i\vec{k}(\vec{x}-\vec{y})}\sum_{\lambda=\pm}\epsilon_i(\vec{k},\lambda)\epsilon^*_j(\vec{k},\lambda)\left(\mu_{k\lambda}\mu'^*_{k\lambda}-\mu^*_{k\lambda}\mu'_{k\lambda}\right).
}
It is easy to check that this quantity is time independent because the term in parenthesis is the Wroskian, $W(\mu_{k\lambda},\mu^*_{k\lambda})$
\eq{
&W(\mu_{k\lambda},\mu^*_{k\lambda})\equiv
\mu_{k\lambda}\mu'^*_{k\lambda}-\mu^*_{k\lambda}\mu'_{k\lambda}\rightarrow \nn \\&W'(\mu_{k\lambda},\mu^*_{k\lambda})=\cancel{\mu'_{k\lambda}\mu'^*_{k\lambda}}+\mu_{k\lambda}\mu''^*_{k\lambda}-\cancel{\mu'^*_{k\lambda}\mu'_{k\lambda}}-\mu^*_{k\lambda}\mu''_{k\lambda}=-\omega^2_{k\lambda}\lvert\mu_{k\lambda}\rvert^2+\omega^2_{k\lambda}\lvert\mu_{k\lambda}\rvert^2=0.
}
Hence, we can compute the commutator at the temporal infinity when $a\gg a_{\rm osc}$ such that $\theta'\rightarrow 0$ and $\omega_{k\lambda}$ does not depend on $\lambda$ (see eqs.~(\ref{alpvelocity}) and (\ref{omega})) so neither does $\mu_{k\lambda}$. Hence it can be factor out from the sum over polarizations
\begin{eqnarray}\label{result}
 \left[A_i(\vec{x}),\Pi_j(\vec{y})\right]_{\infty}&=&\int\frac{d^3k}{(2\pi)^3}e^{i\vec{k}(\vec{x}-\vec{y})}W(\mu_{k\lambda}\mu^*_{k\lambda})_{\infty}\sum_{\lambda=\pm}\epsilon_i(\vec{k},\lambda)\epsilon^*_j(\vec{k},\lambda) \nonumber \\
 &=&\int\frac{d^3k}{(2\pi)^3}e^{i\vec{k}(\vec{x}-\vec{y})}W(\mu_{k\lambda}\mu^*_{k\lambda})_{\infty}\left(\delta_{ij}-\frac{k_ik_j}{k^2}\right),   
\end{eqnarray}
where the subscript $\infty$ means that the quantities are evaluated at the temporal infinity. Thus comparing eqs.~(\ref{canonicalcommutator}) and (\ref{result}) we obtain
\eq{\label{wroskiancondition}
W(\mu_{k\lambda}\mu^*_{k\lambda})=i,
}
where we used that the Wroskian is time independent such that it can be computed at any time.

The second initial condition comes from considerations about the Hamiltonian. This is calculated as follows
\eq{
\mathcal{H}_\gamma=A'_i\Pi_i-\lag_\gamma=\frac{1}{2}\Pi_i^2+\frac{1}{2}\left[\left(\partial_i A_j\right)^2-\left(\partial_i A_i\right)^2\right]-\frac{1}{2}g_{\phi\rm DR}\phi'\epsilon^{ijk}A_i\partial_jA_k.
}
Now using the expansion of the photon field and its conjugated momentum in eqs.~(\ref{quantizedphoton}), (\ref{conjugatedmomentum}), the commutation relations between creation and annihilation operators in eq.~(\ref{creationandannihilationoperators}) and the properties of the polarization vectors in eq.~(\ref{polarizationvectorrelation}), we obtain
\eq{\label{hamiltonian}
H=\int d^3x~\mathcal{H}_\gamma &=\frac{1}{2} \int\frac{d^3k}{(2\pi)^3}\sum_{\lambda=\pm}\left(\lvert\mu'_{k\lambda}\rvert^2+\omega^2_{k\lambda}\lvert\mu_{k\lambda}\rvert^2\right)(a^{\dagger}_{\vec{k}\lambda}a_{\vec{k}\lambda}+a_{\vec{k}\lambda}a^{\dagger}_{\vec{k}\lambda})\nn\\
&+\frac{1}{2}\int\frac{d^3k}{(2\pi)^3}\sum_{\lambda=\pm}\left(\mu'^2_{k\lambda}+\omega^2_{k\lambda}\mu^2_{k\lambda}\right)a_{\vec{k}\lambda}a_{-\vec{k}\lambda}+\hc\equiv \int\frac{d^3k}{(2\pi)^3}\sum_{\lambda=\pm} h_{k\lambda},
}
where $h_{k\lambda}$ has units of $E^{-2}$. The Hamiltonian is time-dependent and can only be diagonalized at a fixed time $\eta_0$. This can be done just by imposing the condition\footnote{This condition is always fulfilled in Minkowski space-time without background fields where the mode functions are $\mu_k=\frac{1}{\sqrt{2\omega_k}}e^{-i\omega_k\eta}$ such that the Hamilitonian is diagonal for any time.}
\eq{\label{instantaneousdiagonalization}
\mu'_{k\lambda}(\eta_0)=-i\omega_{k\lambda}(\eta_0)\mu_{k\lambda}(\eta_0),
}
providing the second initial condition for the mode functions. 

Finally, in the Heisenberg picture (the operators evolve in time but the quantum states do not evolve) we start with the vacuum state $\rvert 0\rangle$ (zero particles) at $\eta_0$. Thus the \textit{single particle occupation number} $n_{k\lambda}$ is defined from the Hamiltonian as\footnote{Notice that in QFT with Hamiltonian 
$h_{k}=\frac{1}{2}\omega_k(a^{\dagger}_k a_k+a_k a^{\dagger}_k)$, with $\omega_{k}$ time independent such that there is no cross terms in the Hamiltonian, and multiparticle states $\rvert p_1...p_n\rangle$, the expectation value of the Hamiltonian is
\eq{
\frac{\langle p_1...p_n\lvert h_k\rvert p_1...p_n\rangle}{\langle p_1...p_n\lvert p_1...p_n\rangle}=\omega_{k\lambda}\frac{\langle p_1...p_n\lvert a^{\dagger}_{k\lambda}a_{k\lambda}+\frac{1}{2}V\rvert p_1...p_n \rangle}{\langle p_1...p_n\lvert p_1...p_n\rangle}=\omega_{k\lambda} V\left(\frac{\langle p_1...p_n\lvert a^{\dagger}_{k\lambda}a_{k\lambda}\rvert p_1...p_n \rangle}{V\langle p_1...p_n\lvert p_1...p_n\rangle}+\frac{1}{2}\right)\equiv \omega_{k\lambda} V\left(n_{k\lambda}+\frac{1}{2}\right),\nn
}
where in the last step we have defined the dimensionless quantity
\eq{\nn
n_{k\lambda}=\frac{\langle p_1...p_n\lvert a^{\dagger}_{k\lambda}a_{k\lambda}\rvert p_1...p_n \rangle}{V\langle p_1...p_n\lvert p_1...p_n\rangle},
}
that counts how many particles with momentum $\vec{k}$ are there in the corresponding multiparticle state.
Our case is analogous but the responsible of a non-vanishing $n_{k\lambda}$ is the time dependent Hamiltonian, that makes $\rvert 0 \rangle$ to be an eigenstate with zero particles only at $\eta_0$.
}
\eq{\label{singleparticle}
\langle 0\lvert h_{k\lambda} \rvert 0\rangle\equiv\omega_{k\lambda}\left(n_{k\lambda}+\frac{1}{2}\right)V,
}
where $V$ is the (infinite) volume and $n_{k\lambda}$ is the number of particles with momentum between $\vec{k}$ and $\vec{k} + d\vec{k}$ per polarization and per unit volume 
\eq{
n_{k\lambda}=\frac{d n_\lambda}{d^3x d^3k}.
}
This definition tells us that the non diagonal form of the Hamiltonian for times $n>\eta_0$ can be interpreted in terms of creation of particles from the vacuum. Mathematically this is because the state $\rvert 0 \rangle$ is only an eigenstate of the hamiltonian at the time $\eta_0$. On the other hand, we can evaluate the left-hand side to obtain
\eq{
\langle 0\lvert h_{k\lambda} \rvert 0\rangle=\frac{1}{2}\left(\lvert\mu'_{k\lambda}\rvert^2+\omega^2_{k\lambda}\lvert\mu_{k\lambda}\rvert^2\right)V,
}
where we used 
\eq{
\langle 0\lvert a_{k\lambda}a^{\dagger}_{k\lambda}\rvert 0 \rangle = \langle 0\lvert a^{\dagger}_{k\lambda} a_{k\lambda} + V \rvert 0 \rangle = V\langle 0\lvert 0 \rangle = V, \textrm{ with }\langle 0\lvert 0\rangle =1.
}
Hence
\eq{\label{occupationnumber}
n_{k\lambda}=\frac{\lvert\mu'_{k\lambda}\rvert^2+\omega^2_{k\lambda}\lvert\mu_{k\lambda}\rvert^2}{2\omega_{k\lambda}}-\frac{1}{2}.
}
In other words, initially in the vacuum state in Heisenberg picture, there are no particles due to the condition in eq.~(\ref{instantaneousdiagonalization}). However, the background field creates particles out of the vacuum with energy given by eq.~(\ref{singleparticle}).

\subsubsection{WKB approach}
In this work we will use the adiabatic WKB representation to solve the eom (\ref{modefunctioneom})
\eq{\label{adianaticrep1}
\mu_{k\lambda}\equiv\frac{\alpha_{k\lambda}(\eta)}{\sqrt{2\omega_{k\lambda}(\eta)}}e^{-i\Psi_{k\lambda}(\eta)}+\frac{\beta_{k\lambda}(\eta)}{\sqrt{2\omega_{k\lambda}(\eta)}}e^{i\Psi_{k\lambda}(\eta)},
}
where $\alpha_{k\lambda}$, $\beta_{k\lambda}$ are the \textit{Bogoliubov} coefficients and $\Psi_{k\lambda}$ is the accumulated phase
\eq{
\Psi_{k\lambda}(\eta)\equiv\int_{\eta_0}^{\eta}d\eta'\omega_{k\lambda}(\eta').
}
The time derivative is taken as if the Bogoliubov coefficients and $\omega_{k\lambda}$ were time independent
\eq{\label{adiabaticrep2}
\mu'_{k\lambda}\equiv-i\alpha_{k\lambda}(\eta)\sqrt{\frac{\omega_{k\lambda}(\eta)}{2}}e^{-i\Psi_{k\lambda}(\eta)}+i\beta_{k\lambda}(\eta)\sqrt{\frac{\omega_{k\lambda}(\eta)}{2}}e^{i\Psi_{k\lambda}(\eta)}.
}
This condition implies non-trivial relations among the Bogoliubov coefficients
\eq{
\alpha'_{k\lambda}(\eta)=\frac{\omega'_{k\lambda}(\eta)}{2\omega_{k\lambda}(\eta)}\beta_{k\lambda}(\eta)e^{2i\Psi_{k\lambda}(\eta)},\quad \beta'_{k\lambda}(\eta)=\frac{\omega'_{k\lambda}(\eta)}{2\omega_{k\lambda}(\eta)}\alpha_{k\lambda}(\eta)e^{-2i\Psi_{k\lambda}(\eta)}.
}
It is easy to check that the WKB mode functions are then solutions of eq.~(\ref{modefunctioneom}). The Wroskian condition in eq.~(\ref{wroskiancondition}) imposes the following normalization of the Bogoliubov coefficients
\eq{\label{bogoliubovnormalization}
\lvert \alpha_{k\lambda}\rvert^2-\lvert \beta_{k\lambda}\rvert^2=1.
}
On the other hand, combining the instantaneous diagonalization of the Hamiltonian in eq.~(\ref{instantaneousdiagonalization}) and the normalization of the Bogoliubov coefficients one arrives to the initial conditions for the Bogoliubov coefficients
\eq{
\alpha_{k\lambda}(\eta_0)=1,\quad \beta_{k\lambda}(\eta_0)=0.
}
On the mode functions this translates into
\eq{\label{modefunctioninitialconditions}
\mu_{k\lambda}(\eta_0)=\frac{1}{\sqrt{2\omega_{k\lambda}(\eta_0)}},\quad \mu'_{k\lambda}(\eta_0)=-i\sqrt{\frac{\omega_{k\lambda}(\eta_0)}{2}}.
}

Using the expression for the single particle occupation number in eq.~(\ref{occupationnumber}) and the adiabatic representation for the mode functions in eqs.~(\ref{adianaticrep1}) and (\ref{adiabaticrep2}) we obtain
\eq{
n_{k\lambda}=\frac{\omega_{k\lambda}\lvert\beta_{k\lambda}\rvert^2}{\omega_{k\lambda}}=\lvert\beta_{k\lambda}\rvert^2,
}
meaning that particle production only depends on the Bogoliubov coefficient $\beta_{k\lambda}$ that parametrizes the departure of the mode function from the positive frequency mode. Since we expect small particle occupation numbers, one can do the following approximations
\eq{\label{betaeq}
\alpha_{k\lambda}\approx 1,\quad \beta'_{k\lambda}(\eta)=\frac{\omega'_{k\lambda}(\eta)}{2\omega_{k\lambda}(\eta)}e^{-2i\Psi_{k\lambda}(\eta)},
}
and we only need to compute $\beta_{k\lambda}$.
Eq.~(\ref{betaeq}) will be used to obtain an analytic insight of the dark photon production. 

\subsubsection{Mathieu analysis}
Let us carefully inspect the eom for the mode functions in eq.~(\ref{modefunctioneom}) when the ALP is already oscillating around one of the minima of its potential and the coupling $g_{\phi\rm DR}$ is activated from the moment $\eta_{\rm DR}$ when the scale factor is $a_{\rm DR}$. From  eq.~(\ref{alpvelocity}) and neglecting $H/m_\phi$ we have 
\eq{
\theta'\approx \delta\theta_{\rm DR}a^{\frac{3}{2}}_{\rm DR}a(\eta)^{-\frac{1}{2}}m_\phi\cos(m_\phi t(\eta)+\nu),
}
where $\delta_{\rm DR}$ is the maximum displacement from the minimum of the potential at $\eta_{\rm DR}$. The frequency 
in eq.~(\ref{omega}) is thus given by
\eq{
\omega^2_{k\lambda}=k^2-k\lambda g_{\phi\rm DR}f_\phi \delta\theta_{\rm DR}a^{\frac{3}{2}}_{\rm DR}m_\phi a(\eta)^{-\frac{1}{2}}\cos(m_\phi t(\eta)+\nu).
}
To convert eq.~(\ref{modefunctioneom}) in a Mathieu equation
\eq{\label{generalmathieu}
\frac{d^2y}{dx^2}+\left[p-2q\cos(2x)\right]y=0
}
we come back to comoving time obtaining 
\eq{
\ddot{\mu}_{k\lambda}+\dot{\mu}_{k\lambda}H+\frac{k^2-k\lambda g_{\phi\rm DR}f_\phi \delta\theta_{\rm DR}a^{\frac{3}{2}}_{\rm DR}m_\phi a(t)^{-\frac{1}{2}}\cos(m_\phi t+\nu)}{a(t)^2}\mu_{k\lambda}=0.
}
Now we define a new variable $z\equiv \frac{1}{2}(m_\phi t + \nu)$, such that
\eq{
\frac{d^2\mu_{k\lambda}}{dz^2}+2\frac{H}{m_\phi}\frac{d\mu_{k\lambda}}{dz}+4\frac{k^2-k\lambda g_{\phi\rm DR}f_\phi \delta\theta_{\rm DR}a^{\frac{3}{2}}_{\rm DR}m_\phi a(z)^{-\frac{1}{2}}\cos(2z)}{m_\phi^2 a(z)^2}\mu_{k\lambda}=0.
}
Neglecting $H/m_\phi$ during oscillations and rearranging the last term we finally arrive to the result
\eq{\label{mymathieu}
\frac{d^2\mu_{k\lambda}}{dz^2}+\left[\frac{4k^2}{m_\phi^2 a(z)^2}-2\lambda g_{\phi\rm DR}f_\phi\delta\theta_{\rm DR}\frac{2k}{m_\phi a(z)}\left(\frac{a_{\rm DR}}{a(z)}\right)^{\frac{3}{2}}\cos(2z)\right]\mu_{k\lambda}=0.
}
Comparing eqs.~(\ref{generalmathieu}) and (\ref{mymathieu}) we find
\eq{
p=\frac{4k^2}{m_\phi^2 a^2},\quad q=\lambda g_{\phi\rm DR}f_\phi\delta\theta_{\rm DR}\frac{2k}{m_\phi a}\left(\frac{a_{\rm DR}}{a}\right)^{\frac{3}{2}}.
}
The Mathieu equation is known to possess instability bands for certain values of $p$ and $q$ where the solution has an exponential growth. For $q\gg 1$ a large region of the parameter space is unstable, leading to broad parametric resonance. Nevertheless in this work we will focus in the narrow resonance regime, meaning  $q\ll 1$. In this case, parametric amplification occurs in the first resonance band given by $1-\lvert q\rvert <p < 1+\lvert q\rvert$
\eq{
1- g_{\phi\rm DR}f_\phi\delta\theta_{\rm DR}\frac{2k}{m_\phi a}\left(\frac{a_{\rm DR}}{a}\right)^{\frac{3}{2}}<\frac{4k^2}{m_\phi^2 a^2}<1+g_{\phi\rm DR}f_\phi\delta\theta_{\rm DR}\frac{2k}{m_\phi a}\left(\frac{a_{\rm DR}}{a}\right)^{\frac{3}{2}}.
}
Taking square root and using that $q$ is small
\eq{
1- g_{\phi\rm DR}f_\phi\delta\theta_{\rm DR}\frac{k}{m_\phi a}\left(\frac{a_{\rm DR}}{a}\right)^{\frac{3}{2}}<\frac{2k}{m_\phi a}<1+ g_{\phi\rm DR}f_\phi\delta\theta_{\rm DR}\frac{k}{m_\phi a}\left(\frac{a_{\rm DR}}{a}\right)^{\frac{3}{2}}.
}
In view of the previous result and using again that $q$ is small, one can substitute $\frac{k}{m_\phi a}=\frac{1}{2}$ and then
\eq{
1-\frac{1}{2}g_{\phi\rm DR}f_\phi\delta\theta_{\rm DR}\left(\frac{a_{\rm DR}}{a}\right)^{\frac{3}{2}}<\frac{2k}{m_\phi a}<1+\frac{1}{2}g_{\phi\rm DR}f_\phi\delta\theta_{\rm DR}\left(\frac{a_{\rm DR}}{a}\right)^{\frac{3}{2}}. 
}
Hence, the condition for small $q$ or, equivalently, narrow resonance is
\eq{\label{narrowresonance}
g_{\phi\rm DR}f_\phi\delta\theta_{\rm DR}\lesssim 1.
}
Consequently, the center of the instability band is at $k/a=m_\phi/2$ and the width is proportional to $g_{\phi\rm DR}f_\phi\delta\theta_{\rm DR}$. Equivalently we can also substitute $a=2k/m_\phi$ to get
\eq{\label{resonantband}
1-\frac{1}{2}g_{\phi\rm DR}f_\phi\delta\theta_{\rm DR}\left(\frac{a_{\rm DR}m_\phi}{2k}\right)^{\frac{3}{2}}<\frac{2k}{m_\phi a}<1+\frac{1}{2}g_{\phi\rm DR}f_\phi\delta\theta_{\rm DR}\left(\frac{a_{\rm DR}m_\phi}{2k}\right)^{\frac{3}{2}}
}

Now we can estimate the enhancement that the mode function experiences due to parametric resonance that will translate into an enhancement on particle production. This can be computed as 
\eq{
\mu_{k\lambda}\propto \exp{\left(\lvert\frac{q}{2}\rvert z_{\rm res}\right)},\quad \textrm{where}\quad   z_{\rm res}=\frac{1}{2}m_\phi(t_{\rm end}-t_{\rm init}),
}
and $t_{\rm init}$, $t_{\rm end}$ are the times in which a mode $k$ enters and leaves the resonant band, respectively. From eq.~(\ref{resonantband}) we can obtain the size of the scale factor at those times
\eq{
a_{\rm init}=\frac{2k}{m_\phi}\left[1-\frac{g_{\phi\rm DR}f_\phi}{2}\delta\theta_{\rm DR}\left(\frac{a_{\rm DR}m_\phi}{2k}\right)^{\frac{3}{2}}\right],\quad a_{\rm end}=\frac{2k}{m_\phi}\left[1+\frac{g_{\phi\rm DR}f_\phi}{2}\delta\theta_{\rm DR}\left(\frac{a_{\rm DR}m_\phi}{2k}\right)^{\frac{3}{2}}\right].
}
Using that the coupling $g_{\phi\rm DR}$ is only active during matter domination epoch and eq.~(\ref{scalefactor}) expressing $t_0$ as a function of the Hubble expansion rate when particle production starts, $H_{\rm DR}\equiv H(a_{\rm DR})$, one obtains
\begin{eqnarray}
  t_{\rm ini}&=&\frac{3}{2H_{\rm DR}}\left(\frac{2k}{a_{\rm DR}m_\phi}\right)^{\frac{3}{2}}\left[1-g_{\phi\rm DR}f_\phi\delta\theta_{\rm DR}\left(\frac{a_{\rm DR}m_\phi}{2k}\right)^{\frac{3}{2}}\right],\nonumber \\
  t_{\rm end}&=&\frac{3}{2H_{\rm DR}}\left(\frac{2k}{a_{\rm DR}m_\phi}\right)^{\frac{3}{2}}\left[1+g_{\phi\rm DR}f_\phi\delta\theta_{\rm DR}\left(\frac{a_{\rm DR}m_\phi}{2k}\right)^{\frac{3}{2}}\right].  
\end{eqnarray}

Hence
\eq{
z_{\rm res}=\frac{9}{4}g_{\phi\rm DR}f_\phi\delta\theta_{\rm DR}\frac{m_\phi}{H_{\rm DR}},
}
and the total growth exponent is given by
\eq{
\frac{\lvert q\rvert}{2} z_{\rm res}=\frac{9}{4}\left(g_{\phi\rm DR}f_\phi\delta\theta_{\rm DR}\right)^2\frac{m_\phi}{H_{\rm DR}}\left(\frac{a_{\rm DR}m_\phi}{2k}\right)^{\frac{3}{2}}.
}
Since for kinematics $m_\phi=2k/a\lesssim 2k/a_{\rm DR}$, the most enhanced mode is $k=a_{\rm DR} m_\phi /2$. The enhancement is proportional to the coupling of the ALP to photons, the misalignment angle when the coupling activates and the ratio between the ALP mass and the Hubble expansion rate at the time when the coupling activates. Notice that the later the coupling activates the more enhanced is the mode function since the enhancement is proportional to $m_\phi/H_{\rm DR}$ and $H$ is decreasing in time ($\dot{H}<0$) in a matter dominated universe. As a result, more dark photons are generated, leading to a reduced amount of dark matter today due to energy conservation. This allows for a higher $H_0$ and a lower $S_8$ compared to the values predicted by $\Lambda\rm CDM$, potentially addressing both tensions simultaneously.

\subsubsection{Analytical calculation}
Here we will obtain analytical expressions for the single particle occupation number and the comoving energy density of dark photons using the WKB approach. Since all the mechanisms to generate DM only defer on the precise instant in which the ALP starts oscillating and those oscillations are the main responsible of particle production, we can obtain general expressions that applies to all mechanisms in matter domination epoch, when the coupling of the dark photons to the ALP is active.  

The starting point is eq.~(\ref{betaeq}) when the ALP oscillates and the coupling $g_{\phi\rm DR}$ is activated. Let us first compute the accumulated phase
\eq{\label{accumulatedphase}
\Psi_{k\lambda}=\int_{\eta_{\rm DR}}^{\eta'}d\eta''\omega_{k\lambda}(\eta'')=\int_{\eta_{\rm DR}}^{\eta'}d\eta''\sqrt{k^2-k\lambda g_{\phi\rm DR}f_\phi\theta'}\approx \int_{\eta_{\rm DR}}^{\eta'}d\eta''k=k\left(\eta'-\eta_{\rm DR}\right),
}
where we have ignored at leading order the contribution proportional to $g_{\phi\rm DR}\delta\theta_{DR}$ coming from $\theta'$ since we work in the narrow resonance regime given by eq.~(\ref{narrowresonance}) and $k/a\approx m_\phi/2$ in the resonant band according to eq.~(\ref{resonantband}), meaning that $q<1$.
On the other hand, we also need to compute the quantity
\eq{
\frac{\omega'_{k\lambda}}{2\omega_{k\lambda}}=\frac{-k\lambda g_{\phi\rm DR}f_\phi\theta''}{4\omega^2_{k\lambda}}=\frac{-k\lambda g_{\phi\rm DR}f_\phi}{4(k^2-k\lambda g_{\phi\rm DR}f_\phi\theta')}\left(-2H a \theta'-m_\phi^2a^2\delta\theta\right)\approx \frac{k\lambda g_{\phi\rm DR}f_\phi m_\phi^2a^2\delta\theta}{4(k^2-k\lambda g_{\phi\rm DR}f_\phi\theta')},
}
where in the second equality we use the ALP eom (\ref{alpeom}) and in the last equality we neglected $H/m_\phi$ during oscillations. Using the explicit forms of $\delta\theta$ and $\theta'$ from eqs.~(\ref{alpposition}) and (\ref{alpvelocity}) neglecting $H/m_\phi$, we get
\eq{
\frac{\omega'_{k\lambda}}{2\omega_{k\lambda}}=\frac{k\lambda g_{\phi\rm DR}f_\phi m_\phi^2a^2}{4\left[k^2-k\lambda g_{\phi\rm DR}f_\phi\delta\theta_{\rm DR}a_{\rm DR}^{\frac{3}{2}}a^{-\frac{1}{2}}m_\phi\cos(m_\phi t(\eta)+\nu)\right]}\delta\theta_{\rm DR}\left(\frac{a_{\rm DR}}{a}\right)^{\frac{3}{2}}\sin(m_\phi t(\eta)+\nu).
}
Again, since in the resonant band $k/a\approx m_\phi/2$ and in the narrow resonance regime we have $g_{\phi\rm DR}\delta\theta_{\rm DR}<1$, we can neglect the second term of the denominator and obtain
\eq{
\frac{\omega'_{k\lambda}}{2\omega_{k\lambda}}=\frac{\lambda g_{\phi\rm DR}f_\phi m_\phi^2a^2}{4k}\delta\theta_{\rm DR}\left(\frac{a_{\rm DR}}{a}\right)^{\frac{3}{2}}\sin(m_\phi t(\eta)+\nu).
}
Substituting this quantity and the accumulated phase (\ref{accumulatedphase}) into eq.~(\ref{betaeq}) we obtain
\eq{
\beta_{k\lambda}=\frac{\lambda g_{\phi\rm DR}f_\phi m_\phi^2}{8ik}\delta\theta_{\rm DR}a_{\rm DR}^{\frac{3}{2}}\int_{\eta_{\rm DR}}^{\eta}d\eta'a(\eta')^{\frac{1}{2}}\left(e^{i\psi_{k\lambda}^{-}(\eta')}-e^{-i\psi_{k\lambda}^{+}(\eta')}\right),
}
where we have written $\sin(m_\phi t +\nu)=\frac{e^{i(m_\phi t +\nu)}-e^{-i(m_\phi t +\nu)}}{2i}$ and the phases are defined as
\eq{
\psi_{k\lambda}^{\pm}(\eta')=\pm 2k(\eta'-\eta_{\rm DR})+m_\phi t(\eta')+\nu.
}
The integrand oscillates quickly so we can apply the stationary phase principle. This establishes that an oscillatory integral of the form
\eq{
\int_{a}^b dx~g(x)e^{i f(x)}
}
where $f(x)$ and $g(x)$ are not oscillatory, is dominated by the points $x_0$ where the phase is stationary
\eq{
\left.\frac{d f}{dx}\right|_{x_0}=0, \quad x_0\in[a,b], \quad \left.\frac{d^2 f}{dx^2}\right|_{x_0}\neq 0
}
and the result of the integral is given, at leading order by
\eq{
\int_{a}^b dx~g(x)e^{i f(x)}= \sum_{x_0}g(x_0)e^{i f(x_0)+\textrm{sign}(f''(x_0))i\frac{\pi}{4}}\left(\frac{2\pi}{\lvert f''(x_0)\rvert}\right)^{\frac{1}{2}}.
}

Applying the stationary phase principle to our case, the integral is dominated by the instants when
\eq{
\left.\frac{d\psi^{-}_{k\lambda}}{d\eta}\right|_{\eta_k}=-2k+m_\phi a(\eta_k)=0\rightarrow a(\eta_k)=\frac{2k}{m_\phi},\quad \left.\frac{d^2\psi^{-}_{k\lambda}}{d\eta^2}\right|_{\eta_k}=m_\phi a'(\eta_k)=m_\phi H(\eta_k)a(\eta_k)^2,
}
and the result is given by
\eq{
\beta_{k\lambda}=\frac{\lambda g_{\phi\rm DR}f_\phi m_\phi^2}{8ik}\delta\theta_{\rm DR}a_{\rm DR}^{\frac{3}{2}}a(\eta_k)^{\frac{1}{2}}e^{i k(\eta_k-\eta_{\rm DR})+i\frac{\pi}{4}}\left(\frac{2\pi}{m_\phi H(\eta_k)a(\eta_k)^2}\right)^{\frac{1}{2}}.
}
Thus, the single particle occupation number reads 
\eq{
n_{k\lambda}=\lvert\beta_{k\lambda}\rvert^2=\pi\frac{g^2_{\phi\gamma\gamma}f^2_\phi\delta\theta_{\rm DR}^2}{32k^2}\frac{a^3_{\rm DR}m_\phi^3}{H(\eta_k)a(\eta_k)}=\pi\frac{g^2_{\phi\gamma\gamma}f^2_\phi\delta\theta_{\rm DR}^2}{32k^2}\frac{a^3_{\rm DR}m_\phi^3}{H(\eta_k)a(\eta_k)}.
}
Notice that the resulting single particle occupation number is independent of the polarization of the dark photon. Using that $a(\eta_k)=\frac{2k}{m_\phi}$ and eq.~(\ref{Hubble}) we have 
\eq{
H(\eta_k)=H_{\rm DR}\left(\frac{a_{\rm DR} m_\phi}{2k}\right)^{\frac{3}{2}}
}
where $H_{\rm DR} = H_0 a_{\rm DR}^{-\frac{3}{2}}$ and thus
\eq{
n_{k\lambda}=\frac{\pi}{8}\left(g_{\phi\gamma\gamma}f_\phi\delta\theta_{\rm DR}\right)^2 \frac{m_\phi}{H_{\rm DR}}\left(\frac{a_{\rm DR}m_\phi}{2k}\right)^{\frac{3}{2}}.
}

This expression confirms what we already knew from the Mathieu analysis: the mode that experiences more particle creation is $k=a_{\rm DR}m_\phi/2$ and the number of particles created decreases with $k$, the number of particles produced is proportional to the squared of $g_{\phi\rm DR}f_\phi\delta\theta_{\rm DR}$ and the later the coupling $g_{\phi\rm DR}$ is activated, the larger the ratio $m_{\phi}/H_{\rm DR}$ and the most particles are created for a given mode.

Now we obtain the expression for the co-moving energy density of dark photons. This is defined as follows
\eq{
a^4\rho_{\rm DR}\equiv \sum_\lambda\int\frac{d^3k}{(2\pi)^3}\omega_{k\lambda}n_{k\lambda}.
}
At leading order, the frequency $\omega_{k\lambda}\approx k$ is independent of the polarization and so is $n_{k\lambda}$. Hence
\begin{eqnarray}
    a^4\rho_{\rm DR} &=& 2\int_0^\infty \frac{4\pi}{(2\pi)^3}dk~k^2~k~ \frac{\pi}{8}\left(g_{\phi\gamma\gamma}f_\phi\delta\theta_{\rm DR}\right)^2 \frac{m_\phi}{H_{\rm DR}}\left(\frac{a_{\rm DR}m_\phi}{2k}\right)^{\frac{3}{2}} \nonumber \\
    &=& \frac{\left(g_{\phi\rm DR}f_\phi\delta\theta_{\rm DR}\right)^2}{8\pi}\frac{m_\phi}{H_{\rm DR}}\left(\frac{a_{\rm DR}m_\phi}{2}\right)^{\frac{3}{2}}\int_0^\infty dk~k^{\frac{3}{2}}.
\end{eqnarray}

At first glance the integral is divergent. However, at a fixed conformal time $\eta$ the mode that is being produced satisfies $k=m_\phi a(\eta)/2$ according to kinematics and the stationary phase principle, fixing the upper limit of the integral. On the other hand, since dark photons start to be produced at $\eta_{\rm DR}$, the lower limit of the integral is fixed to $k=m_\phi a_{\rm DR}/2$. Hence, the integral is finite giving
\eq{\label{photonenergydensity}
\rho_{\rm DR}=\frac{\left(g_{\phi\rm DR}f_\phi\delta\theta_{\rm DR}\right)^2 m_\phi^4}{320\pi}\frac{m_\phi}{H_{\rm DR}}\left(\frac{a_{\rm DR}}{a}\right)^{\frac{3}{2}}\left[1-\left(\frac{a_{\rm DR}}{a}\right)^{\frac{5}{2}}\right],\quad a>a_{\rm DR}
}
and $\rho_{\rm DR}=0$ for $a<a_{\rm DR}$. Notice that the energy density of dark photons decays slower than that of radiation $\rho_{\rm R}\sim a^{-4}$ due to the continuous support of the background ALP. This expression holds for any mechanism of ALP DM at late times.

Hitherto everything was computed assuming that the ALP is not affected by the production of photons. At leading order in the photon-ALP coupling $g_{\phi\rm DR}$ this is a good approximation for the energy density of dark photons. To include the backreaction on the ALP field for $a>a_{\rm DR}$ we use the energy conservation. 

Using that the ALP energy density scales as $\rho_\phi\sim a^{-3}$ according to eq.~(\ref{ALPenergydensity}), energy conservation demands
\eq{\label{energyconserv}
a^{-3}\frac{d}{dt}\left(a^3\rho_\phi\right)=-a^{-4}\frac{d}{dt}\left(a^4\rho_{\rm DR}\right)\rightarrow \frac{d}{dt}\left(a^3\rho_\phi\right) = -a^{-1} \frac{d}{dt}\left(a^4\rho_{\rm DR}\right).
}
Let us compute the r.h.s of eq.~(\ref{energyconserv}). From eq.~(\ref{photonenergydensity}) we get
\eq{
\frac{d}{dt}\left(a^3\rho_{\phi}\right)=-\frac{(g_{\phi\rm DR}f_\phi\delta\theta_{\rm DR})^2}{128\pi}\frac{a^{\frac{3}{2}}_{\rm DR}m_\phi^5}{H_{\rm DR}}a^{\frac{1}{2}}\dot{a}.
}
Using $\dot{a} = da/dt$ and integrating from $a_{\rm DR}$ we obtain
\eq{
\rho_\phi(a)=\rho_\phi(a_{\rm DR})\left(\frac{a_{\rm DR}}{a}\right)^3-\frac{(g_{\phi\rm DR}f_\phi\delta\theta_{\rm DR})^2}{192\pi}\frac{m_\phi^5}{H_{\rm DR}}\left(\frac{a_{\rm DR}}{a}\right)^{\frac{3}{2}}\left[1-\left(\frac{a_{\rm DR}}{a}\right)^{\frac{3}{2}}\right].
}
We can express $\delta\theta_{\rm DR}$ as a function of the ALP energy density by noticing that when the ALP coupling to DR activates we can safely neglect the ALP kinetic term and its energy density is approximately given by
\eq{\label{deltatheta}
\rho_\phi(a_{\rm DR})\approx \frac{1}{2}m_\phi^2 f_\phi^2 \delta\theta_{\rm DR}^2.
}
The expression for the ALP energy density for $a>a_{\rm DR}$ is thus given by
\begin{eqnarray}
    \label{ALPenergydensity}
\rho_\phi(a)&=&\rho_\phi(a_{\rm DR})\left(\frac{a_{\rm DR}}{a}\right)^3\left[1-\xi\left(a^{\frac{3}{2}}-a_{\rm DR}^{\frac{3}{2}}\right)\right], \nonumber \\
\xi &\equiv& \frac{(g_{\phi\rm DR}f_\phi)^2}{96\pi}\frac{m_\phi^3}{H_{\rm DR} a_{\rm DR}^{\frac{3}{2}}f_\phi^2}=\frac{(g_{\phi\rm DR}f_\phi)^2}{96\pi}\frac{m_\phi^3}{H_0 f_\phi^2},
\end{eqnarray}
where we used $H_{\rm DR}=H_0 a_{\rm DR}^{-\frac{3}{2}}$ in the definition of $\xi$. We assume that $\xi<1$ such that the ALP is a background field and the correction due to the production of dark photons is small. This condition also ensures a positive DM energy density today. 

Hereafter we will follow closely refs.~\cite{DES:2020mpv,McCarthy:2022gok}. One can rewrite this expression in terms of the DM relic density today $\rho^0_{\rm DM}$. For that purpose we set $a=1$ so that the ALP energy density at $a=a_{\rm DR}$ can be written in terms of the today DM relic density as
\eq{\label{rhophi}
\rho_\phi(a_{\rm DR})=\frac{\rho^0_{\rm DM}}{a^3_{\rm DR}}\left[1-\xi\left(1-a_{\rm DR}^{\frac{3}{2}}\right)\right]^{-1},
}
such that
\eq{\label{ALPenergydensity_def}
\rho_\phi(a)=\frac{\rho^0_{\rm DM}}{a^3}\left[1+\xi\frac{1-a^{\frac{3}{2}}}{1-\xi\left(1-a_{\rm DR}^{\frac{3}{2}}\right)}\right],\quad a>a_{\rm DR}
}
expression that holds for $a>a_{\rm DR}$. Applying continuity of the ALP energy density at $a=a_{\rm DR}$ and that it scales as $\rho_{\phi}(a)\sim a^{-3}$ when the coupling to dark photons is not activated, we get
\eq{
\rho_{\phi}(a)=\frac{\rho^0_{\rm DM}}{a^3}\left[1+\xi\frac{1-a^{\frac{3}{2}}_{\rm DR}}{1-\xi\left(1-a^{\frac{3}{2}}_{\rm DR}\right)}\right],\quad a_{\rm osc}<a<a_{\rm DR}.
}
For $a<a_{\rm osc}$, the behavior of the ALP is model dependent as emphasized in sec.~\ref{review}. However, assuming the standard cosmological history of the universe, the ALP energy density will be negligible during most of the radiation domination epoch, independently of the dynamics of the ALP before the onset of the oscillations. Hence, we set $\rho_{\phi}(a) = 0$ for $a<a_{\rm osc}$ without loss of generality.

We can also rewrite the energy density of dark photons in eq.~(\ref{photonenergydensity}) as a function of the today DM energy density and $\xi$
\eq{\label{photonenergydensity_def}
\rho_{\rm DR}=\frac{3}{5}\frac{\rho^0_{\rm DM}}{a^{\frac{3}{2}}}\xi\frac{1-\left(\frac{a_{\rm DR}}{a}\right)^{\frac{5}{2}}}{1-\xi\left(1-a^{\frac{3}{2}}_{\rm DR}\right)}.
}

 On the other hand, imposing that the occupation number of dark photons in eq.~(\ref{occupationnumber}) remains small, $n_{k\lambda}<1$, gives the following relation
\eq{\label{small_number_density}
\xi < \frac{a^{\frac{3}{2}}_{\rm DR}}{24\pi^2\frac{\rho^0_{\rm DM}}{m^4_{\phi}}+a^{\frac{3}{2}}_{\rm DR}(1-a^{\frac{3}{2}}_{\rm DR})},
}
where we used the definition of $\xi$ in eq.~(\ref{ALPenergydensity}) and expressed $\delta\theta_R$ as a function of the today DM using eqs.~(\ref{deltatheta}) and (\ref{rhophi})
Assuming that the corrections to $\Lambda \rm CDM$ will be small, as a first approximation we can use $\Omega_{\rm DM}h^2\approx 0.12$, $h\approx 0.7$ and $H_0\approx 1.5\cdot 10^{-42}$~GeV and the Planck mass $M_p=(8\pi G_N)^{-1/2}=2.43\cdot 10^{18}$~GeV to obtain an estimation of $\rho^0_{\rm DM}$
\eq{
\rho^0_{\rm DM}=\frac{\rho^0_{\rm DM}h^2}{\rho_{\rm crit}}\frac{\rho_{\rm crit}}{h^2}=\Omega_{\rm DM}h^2\frac{3 H_0^2 M^2_p}{h^2}=10^{-47}~\rm{GeV}^4.
}
Substituting we get a lower bound for the ALP mass in the decaying pre-inflationary ALP model
\eq{\label{lowerboundALPmass}
\frac{m_{\phi}}{7~\rm meV}\geq \left[a^{\frac{3}{2}}_{\rm DR}\left(\frac{1}{\xi}-1+a^{\frac{3}{2}}_{\rm DR}\right)\right]^{-\frac{1}{4}}\approx \xi^{\frac{1}{4}}a^{-3/8}_{\rm DR},
}
where in the last step we used that $\xi\ll 1$.

We also need to compute the collision term $\mathcal{Q}$. This is obtained from
\eq{
a^{-3}\frac{d}{dt}\left(a^3 \rho_{\rm DM}\right)=-a^{-4}\frac{d}{dt}\left(a^4 \rho_{\rm DR}\right)=-\mathcal{Q}.
}
Hence
\eq{\label{collisionterm}
\mathcal{Q}=\frac{3}{2}\xi\frac{ H}{a^{\frac{3}{2}} }\frac{\rho^0_{\rm DM}}{1-\xi\left(1-a^{\frac{3}{2}}_{\rm DR}\right)}.
}
Finally we also need $\Gamma$, the inverse lifetime of the decaying DM
\eq{
\Gamma=\frac{\mathcal{Q}}{\rho_{\rm DM\rightarrow \rm DR}},
}
where $\rho_{\rm DM\rightarrow \rm DR}$ is the component of DM that converts into dark radiation. From eq.~(\ref{ALPenergydensity_def})
\eq{
\rho_{\rm DM\rightarrow \rm DR}=\frac{\rho^0_{\rm DM}}{a^3}\xi\frac{1-a^{\frac{3}{2}}}{1-\xi\left(1-a^{\frac{3}{2}}_{\rm DR}\right)}.
}
Thus the inverse lifetime reads
\eq{
\Gamma=\frac{3}{2} H \frac{a^{\frac{3}{2}}}{1-a^{\frac{3}{2}}}.
}
Notice that this expression diverges when $a\rightarrow 1$. This divergence will be regularized numerically as we explain in sec.~\ref{Bayesiananalysis}.

Let us summarize the main results of this section. The ALP energy density as a function of the scale factor is given by
\eq{
\rho_{\phi}(a)=\frac{\rho^0_{\rm DM}}{a^3}\times \begin{cases}
    0 &  a < a_{\rm osc}, \\
   1+\xi\frac{1-a^{\frac{3}{2}}_{\rm DR}}{1-\xi\left(1-a^{\frac{3}{2}}_{\rm DR}\right)} & a_{\rm osc} \leq a \leq a_{\rm DR}, \\
    1+\xi\frac{1-a^{\frac{3}{2}}}{1-\xi\left(1-a^{\frac{3}{2}}_{\rm DR}\right)} & a > a_{\rm DR}.
\end{cases}
}
The DR energy density is
\eq{
\rho_{\rm DR}(a)=\frac{3}{5}\frac{\rho^0_{\rm DM}}{a^{\frac{3}{2}}}\xi\frac{1-\left(\frac{a_{\rm DR}}{a}\right)^{\frac{5}{2}}}{1-\xi\left(1-a^{\frac{3}{2}}_{\rm DR}\right)} \times
\begin{cases}
   0 &  a \leq a_{\rm DR}, \\
    1 & a > a_{\rm DR}.
\end{cases}
}
The inverse lifetime of the decaying ALP is
\eq{
\Gamma=\frac{3}{2}H\frac{a^{\frac{3}{2}}}{1-a^{\frac{3}2}}\times \begin{cases}
   0 &  a \leq a_{\rm DR}, \\
    1 & a > a_{\rm DR}.
\end{cases}
}

\begin{figure}
\centering
\subfigure[]{\includegraphics[scale=0.45]{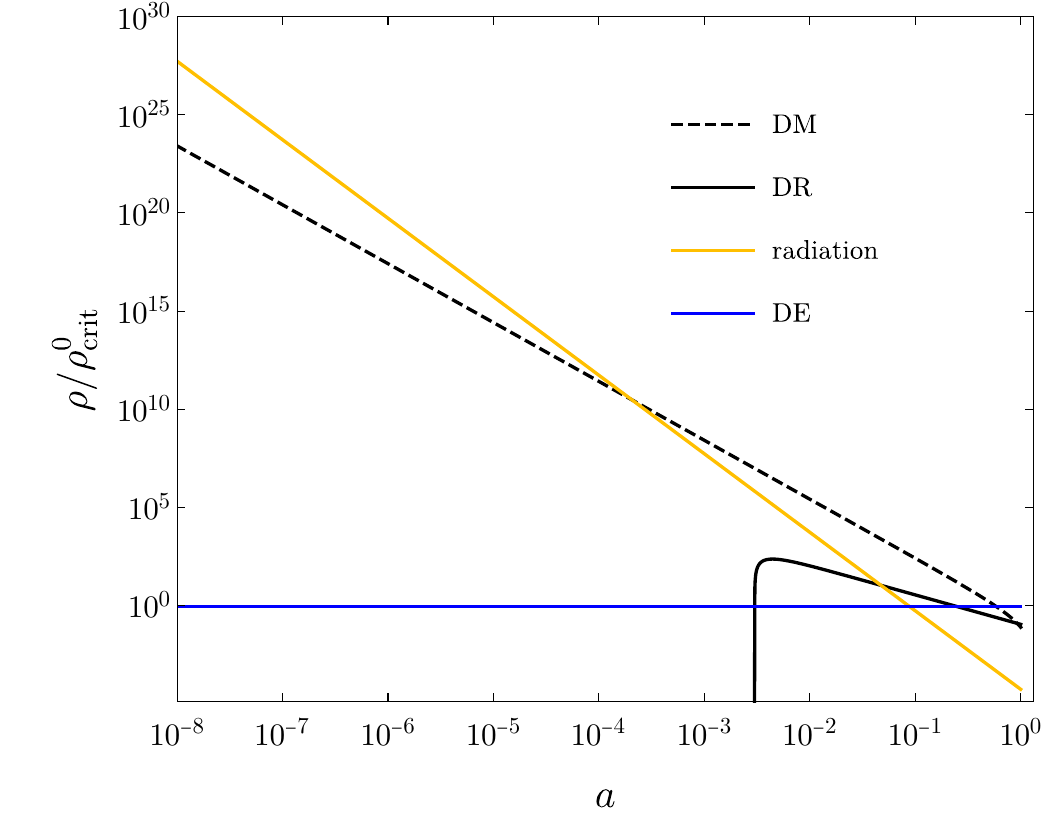}}
\subfigure[]{\includegraphics[scale=0.45]{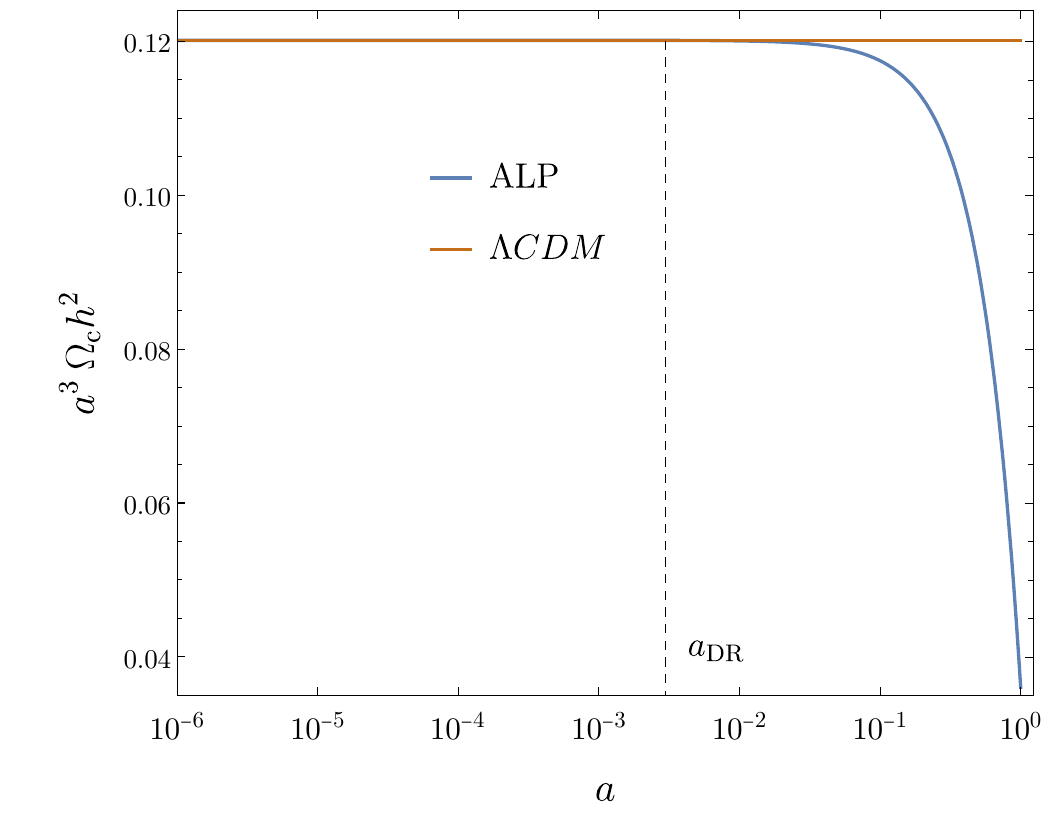}}
\caption{\textit{Left:} energy density of the different species included in our model normalized to the critical energy density today, $\rho^0_{\rm crit}$ for the benchmark point $a_{\rm osc}=10^{-8}$, $a_{\rm DR}=0.003$, $\xi=0.7$. \textit{Right:} for the same benchmark point, comparison between the DM energy density in $\Lambda$CDM and in the ALP model.} 
\label{fig:benchmark_densities}
\end{figure}

The typical behavior of our model compared to $\Lambda$CDM is depicted in figs.~\ref{fig:benchmark_densities} and \ref{fig:benchmark_hubble}. For $\Lambda$CDM we have fixed the background cosmological parameters to the best-fit values from the \textit{Planck} fit to the CMB alone (TT-EE-TE) in ref.~\cite{Planck:2018vyg}($100\theta_s=1.040909$, $100\Omega_b h^2=2.2383 $, $\Omega_c h^2= 0.12011$) where $\theta_s$ is the angular size of the acoustic scale at last scattering (during recombination), $\Omega_b h^2$ is the physical density of baryons today, and $\Omega_c h^2$ is the physical density of CDM today, quantities that are directly constrained by the CMB. 

On the other hand, for the ALP model we need to consider that the CMB constrains the amount of DM at recombination $a=a_{*}\approx 10^{-3}$, when the CMB was released. Hence, our model will satisfy the constraint of the CMB if at $a=a_{*}$ we impose
\eq{\label{ldcd_alp_dm}
\left(\rho^0_{\rm DM}\right)^{\rm ALP}=\left(\rho^0_{\rm DM}\right)^{\rm \Lambda CDM}\left[1+\xi\frac{1-a^{\frac{3}{2}}_{\rm DR}}{1-\xi\left(1-a^{\frac{3}{2}}_{\rm DR}\right)}\right]^{-1},
}
since $a_{\rm osc}<a_{*}<a_{\rm DR}$. 

In fig.~\ref{fig:benchmark_densities} we show the energy densities of the different components of the universe in the ALP model for $a_{\rm osc}=10^{-8}$, $a_{\rm DR}=0.7$ and $\xi=0.003$. In particular, the energy density of the ALP matches that of DM in $\Lambda$CDM until its coupling to DR is activated, as illustrated in the right panel. Later on, part of the ALP energy density converts into DR and the amount of DM dilutes faster than in $\Lambda$CDM. As a consequence, the matter-DE equality is achieved before than in $\Lambda$CDM, allowing for a larger $H_0$ today as shown in the right panel of fig.~\ref{fig:benchmark_hubble}. Notice that the amount of DR is negligible compared to that of DE.

\section{Numerical results}\label{Bayesiananalysis}
All the previous equations are implemented in the software CLASS~\cite{Diego_Blas_2011} using as a reference the modification of CLASS in \url{https://github.com/fmccarthy/class_DMDR} \cite{McCarthy:2022gok} that allows for a decaying DM candidate with a time dependent decay rate. We also use Cobaya~\cite{Torrado:2020dgo} that contains likelihood codes of most recent experiments and interfaces with CLASS for computing the cosmological observables. In the following we explain in detail the modifications introduced in CLASS with respect to \cite{McCarthy:2022gok} and the likelihoods we use in Cobaya to constrain our model.

\subsection{Numerical implementation in CLASS}
To implement the above equations in CLASS we introduce the following changes with respect to \cite{McCarthy:2022gok}
\begin{itemize}
    \item [1.] In the perturbations file, in the Boltzmann equation for the DR perturbations, the DR energy density appears in a denominator leading to huge values for $a\approx a_{\rm DR}$ . Hence we have to distinguish cases. For $a<a_{\rm DR}$ there is no DR, so the perturbations must vanish. For $a>a_{\rm DR}$ we let the original expression but in the background file, we place a ceiling on the DR energy density such that its value can never be less that $10^{-4}$ of its value today.

    \item[2.] Something similar occurs with the inverse lifetime of the decaying ALP before the onset of the oscillations. For $a<a_{\rm DR}$ we put a ceiling on the ratio $\Gamma(a)/H(a)$ such that its value can never be smaller than $10^{-4}$. Hence, the decay rate of the ALP is negligible compared to the expansion of the universe. On the other hand, $\Gamma$ diverges for $a\approx 1$. Thus as in ref.~\cite{McCarthy:2022gok} we place another ceiling on the ratio $\Gamma(a)/H(a)$ such that its value never exceeds 100.
   
\end{itemize}

\subsection{Data}
We constrain the parameters of the ALP decay model and compare its performance against $\Lambda\rm CDM$ using a range of cosmological datasets. These datasets include CMB observations, large-scale structure data (such as CMB lensing and BAO), supernovae luminosity distances (which constrain $H(z)/H_0$), a local measurement of $H_0$, and a direct measurement of $S_8$ obtained from low-redshift galaxy surveys or other probes of the matter power spectrum. The likelihood function is computed using CLASS. Below, we provide a more detailed description of the datasets and likelihoods employed in our analysis.

\subsubsection{\textit{Planck} primary CMB and CMB lensing}

We start by using the \textit{Planck} 2018 likelihood for the primary CMB alone, which includes the low-$\ell$ TT, low-$\ell$ EE, and high-$\ell$ TT/TE/EE (plik) power spectra~\cite{Planck:2019nip}. In addition, we incorporate the 2018 lensing likelihood~\cite{Planck:2018lbu} (clik) that probes the large-scale structure of the late universe.

\subsubsection{Baryon acoustic oscillations}

We include BAO likelihoods from the SDSS DR7~\cite{Howlett:2014opa} and DR16~\cite{eBOSS:2020yzd} surveys. The SDSS DR7 provides a measurement of the BAO scale at $z=0.15$, while DR16 offers measurements at $z=0.38$, $0.51$, $0.7$, and $1.48$.

\subsubsection{Luminosity distances: supernovae}

We use supernovae from the Pantheon Plus sample~\cite{Brout:2022vxf}, which provide luminosity distance measurements over the redshift range $0.001<z<2.26$.

\subsubsection{Cosmic distance ladder: $H_0$ from SH0ES}

We use the likelihood from the SH0ES team that obtained the most recent value of $H_0$: $H_0=73.04\pm1.04 \,\mathrm{km/s/Mpc}$~\cite{Riess:2021jrx}. They calibrate 42 SNe using high-redshift Cepheids and use 277 SNe in the redshift range $0.023 < z < 0.15$ to determine $H_0$. The Cepheids are themselves calibrated using parallax measurements from Cepheids in the Milky Way, as well as those in the Large and Small Magellanic Clouds, NGC 4258 (based on geometric megamaser distance), and M31.

\subsubsection{Matter clustering: $S_8$ from DES}

We use the likelihood of the Dark Energy Survey (DES). DES probes the low-redshift universe using cosmic shear, galaxy clustering data, and their cross-correlation (galaxy-galaxy lensing). The DES-Y3 results~\cite{DES:2021wwk} combine these measurements to yield $S_8 = 0.776 \pm 0.017$.

\subsubsection{Priors on the ALP model parameters}

\begin{figure}[t!]
\centering
\subfigure[]{\includegraphics[scale=0.45]{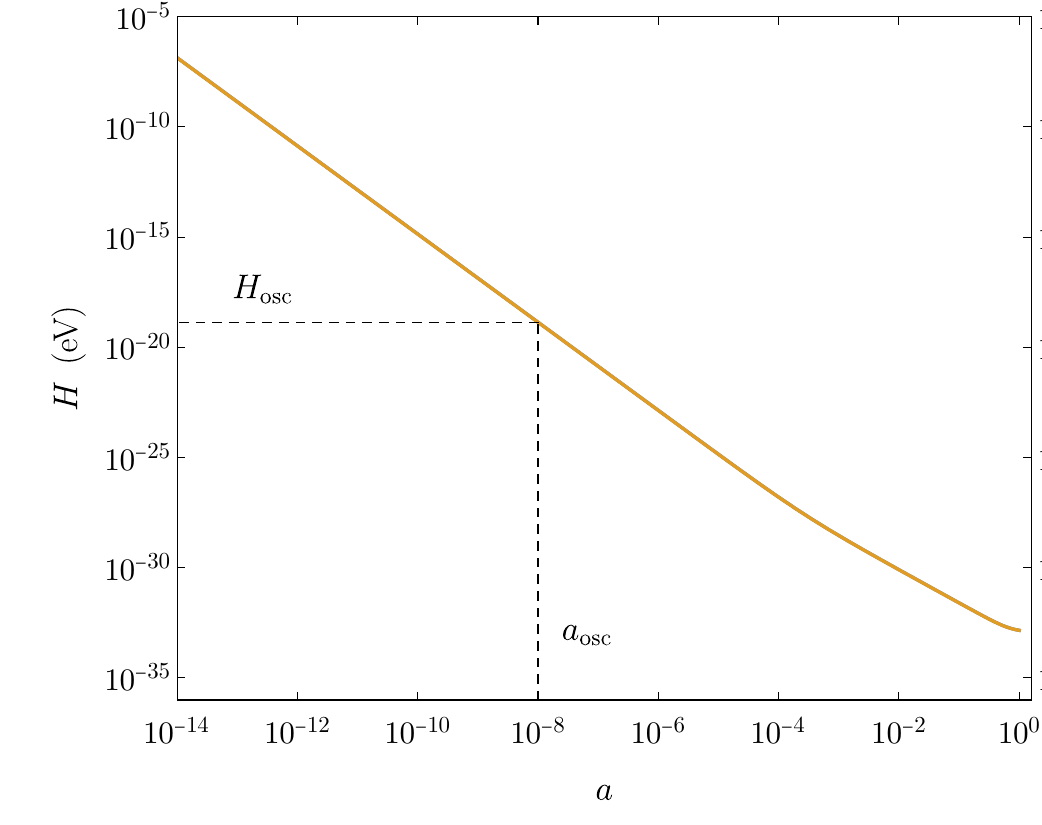}}\hfill
\subfigure[]{\includegraphics[scale=0.45]{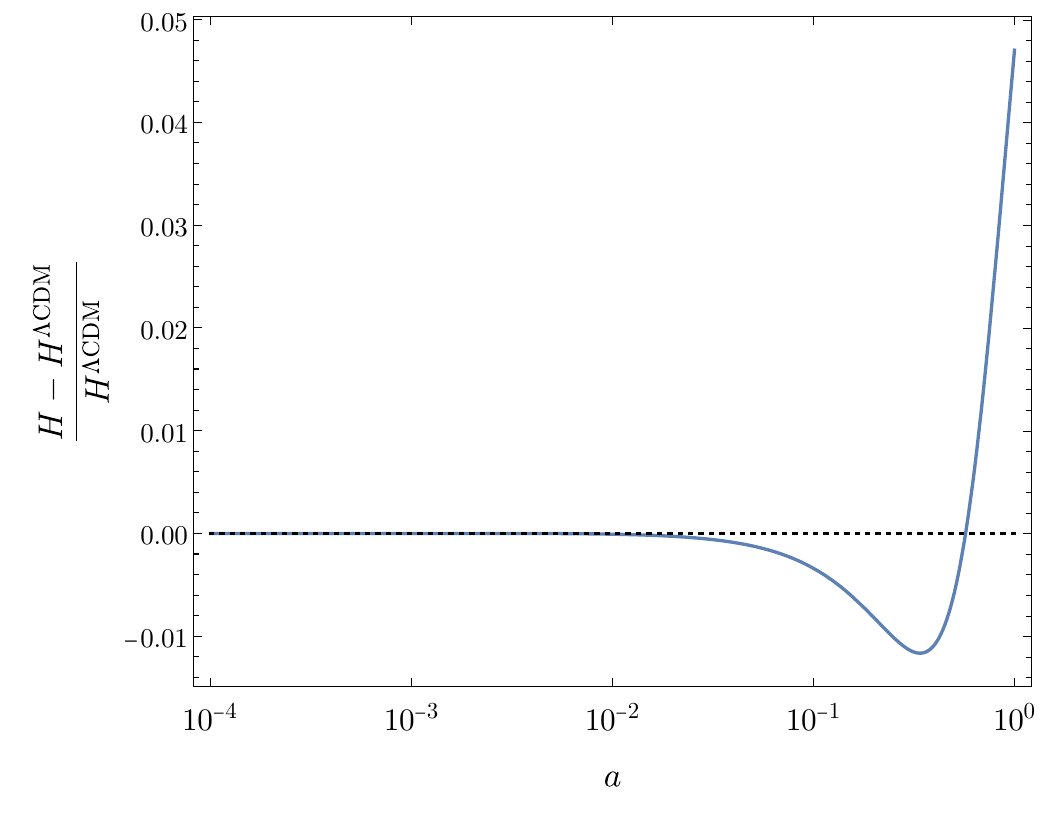}}
\caption{\textit{Left}: Hubble parameter as a function of the scale factor in the ALP model for the benchmark point $a_{\rm osc}=10^{-8}$, $a_{\rm DR}=0.003$, $\xi=0.7$. The rest of the parameter are fixed to their best-fit values from the \textit{Planck} fit to the CMB alone. The DM density is obtained using eq.~(\ref{ldcd_alp_dm}). \textit{Right:} comparison with $\Lambda$CDM. In both cases the Hubble parameter is obtained using the `shooting' method implemented in CLASS.}\label{fig:benchmark_hubble}
\end{figure}

For the sake of simplicity, in this work we fix the parameter $a_{\rm osc}$ that controls the moment in which the ALP oscillations start to behave as DM to be $a_{\rm osc}=10^{-8}$, inside the radiation domination epoch. This choice does not have a significant impact in our conclusions. First of all, because this moment is deeply inside radiation domination epoch when the ALP is just an spectator field. However, it does impose a lower bound on the ALP mass since the oscillation condition is $m_{\phi}>3H_{\rm osc}$. Just by running CLASS using the ALP model, it is easy to check that the Hubble parameter is of size $H_{\rm osc}\sim 10^{-19}~\rm{eV}$ according to fig.~\ref{fig:benchmark_hubble} and then $m_{\phi}>3\cdot 10^{-19}~\rm{eV}$. Later we will show that the condition of small number density of dark photons in eq.~(\ref{small_number_density}) is more stringent. On the other hand, the parameter $\xi$ that governs the ALP to DR conversion needs to be small in order to keep under control the deviations of our model with respect to $\Lambda$CMD and also to be compatible with the approximation of narrow resonance in the Mathieu analysis of the previous section. Hence we will focus in the region $\xi\in \left[10^{-5},10^{0}\right]$. Finally, since we suppose that the ALP to DR conversion occurs during matter domination epoch, we constrain $a_{\rm DR}$ to be in the interval $\left[10^{-3},1\right]$. In view of the previous constrains we use linear priors on $\log_{10}\xi$ and $\log_{10}a_{\rm DR}$.

\subsection{Results}
 \begin{table*}
\resizebox{18cm}{!} { 
\begin{tabular} { |l|  c c |c c| c c|}\hline
&\multicolumn{2}{c|}{\textit{Planck} primary CMB}&\multicolumn{2}{c|}{$+\phi\phi+$BAO+SN+DES}&\multicolumn{2}{c|}{+SH0ES}\\\cline{2-7}
 Parameter &  ALP model&$\Lambda$CDM&  ALP model&$\Lambda$CDM&  ALP model&$\Lambda$CDM\\
\hline\hline

{$\log(10^{10} A_\mathrm{s})$} & $3.0511^{+0.0086}_{-0.014} $& 
$3.052^{+0.010}_{-0.014}            $& $3.048^{+0.014}_{-0.012} $& 
$3.047\pm 0.012            $& $3.049\pm 0.012$&
$3.050^{+0.014}_{-0.012}$\\

{$n_\mathrm{s}   $} & $0.9650\pm 0.0045$& $0.9650\pm 0.0044$& $0.9669\pm 0.0037$& $0.9670\pm 0.0035$& $0.9692\pm 0.0036          $& $0.9696\pm 0.0036$\\

{$100\Omega_\mathrm{b} h^2$} & $2.235\pm 0.015$& $2.236\pm 0.015$& $2.246\pm 0.014          $& $2.245\pm 0.013        $& $2.255\pm 0.013  $& $2.256\pm 0.013        $\\

{$\Omega_\mathrm{c} h^2$} & $0.1195^{+0.0018}_{-0.0014}$& $0.1200\pm 0.0014$& $0.1184^{+0.0012}_{-0.00088}$&     $0.11879\pm 0.00087$& $0.1176^{+0.0010}_{-0.00091}$& $0.11785\pm 0.00086$\\

{$\tau_\mathrm{reio}$} & $< 0.0597 $& $< 0.0600$& $0.0570\pm 0.0063 $& $0.0567\pm 0.0061          $& $0.0586^{+0.0073}_{-0.0056}$ & $0.0589^{+0.0075}_{-0.0055}$\\

{$100\theta_\mathrm{s}$} & $1.04186\pm 0.00029 $& $1.04185\pm 0.00030$& $1.04195\pm 0.00028         $& $1.04195\pm 0.00027$& $1.04206\pm 0.00028$& $1.04207\pm 0.00027        $\\

{$\log_{10}\xi          $}  &
$< -2.76     $&
---                         & 
$< -2.79                 $&
---                         &
$< -2.83                  $&
---                         \\
{$\log_{10}(a_{\rm DR}) $} & 
$\rm{unconstrained}              $&
---                         & 
$\rm{unconstrained}     $& 
---                         &
$\rm{unconstrained} $&
---  \\


\hline\hline

$H_0                       $ & $67.39\pm 0.62    $& $67.35\pm 0.61       $& $67.91\pm 0.40              $& $67.90\pm 0.39     $& $68.36\pm 0.40$& $68.37\pm 0.39              $\\

$\Omega_\mathrm{m}         $ & $0.3138\pm 0.0089$& $0.3153^{+0.0077}_{-0.0087} $& $0.3069\pm 0.0056         $& $0.3078\pm 0.0052 $& $0.3012\pm 0.0052$& $0.3018\pm 0.0050$\\

$S_8                       $& $0.832\pm 0.017$& $0.834\pm 0.016$& $0.818\pm 0.010 $& $0.8187\pm 0.0099$& $0.808\pm 0.010$& $0.8091\pm 0.0097$\\
\hline
\end{tabular}
}
\caption{The 68\% confidence limits on the base $\Lambda$CDM and ALP model parameters, along with the derived parameters $H_0$, $\Omega_m$ and $S_8$, for the ALP model and $\Lambda$CDM analyses.}\label{tab:68_confidence_intervals}
 \end{table*}

\begin{figure}[ht!]
\includegraphics[width=12cm]{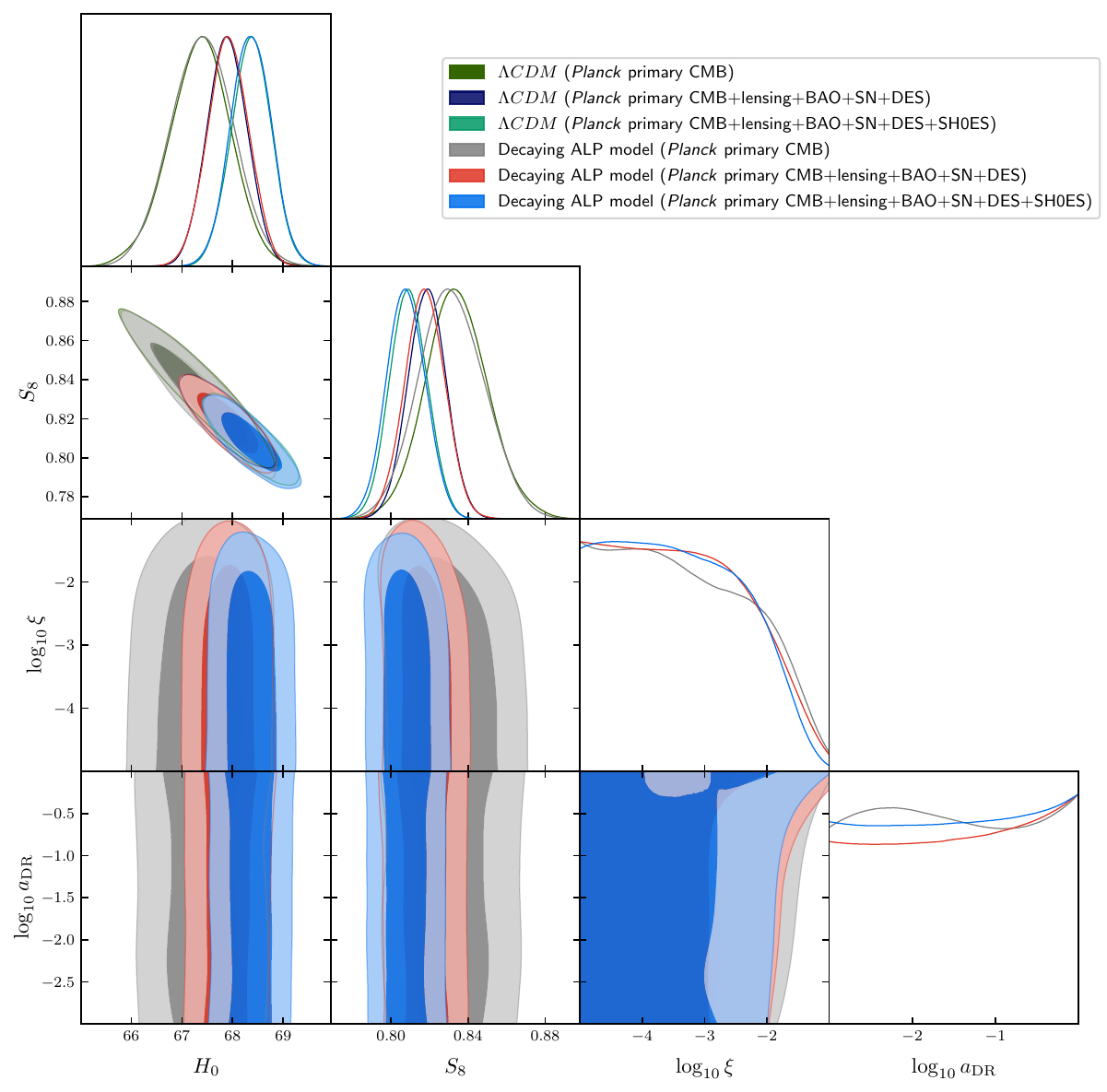}
\centering
\caption{Posteriors on $H_0$ and $S_8$, as well as the $\log_{10}\xi$ and $\log_{10} a_{\rm DM}$ parameters for ALP decaying model analysis.} 
\label{fig:posteriors}
\end{figure}

\begin{figure}[ht!]
\centering
\subfigure[]{\includegraphics[scale=0.40]{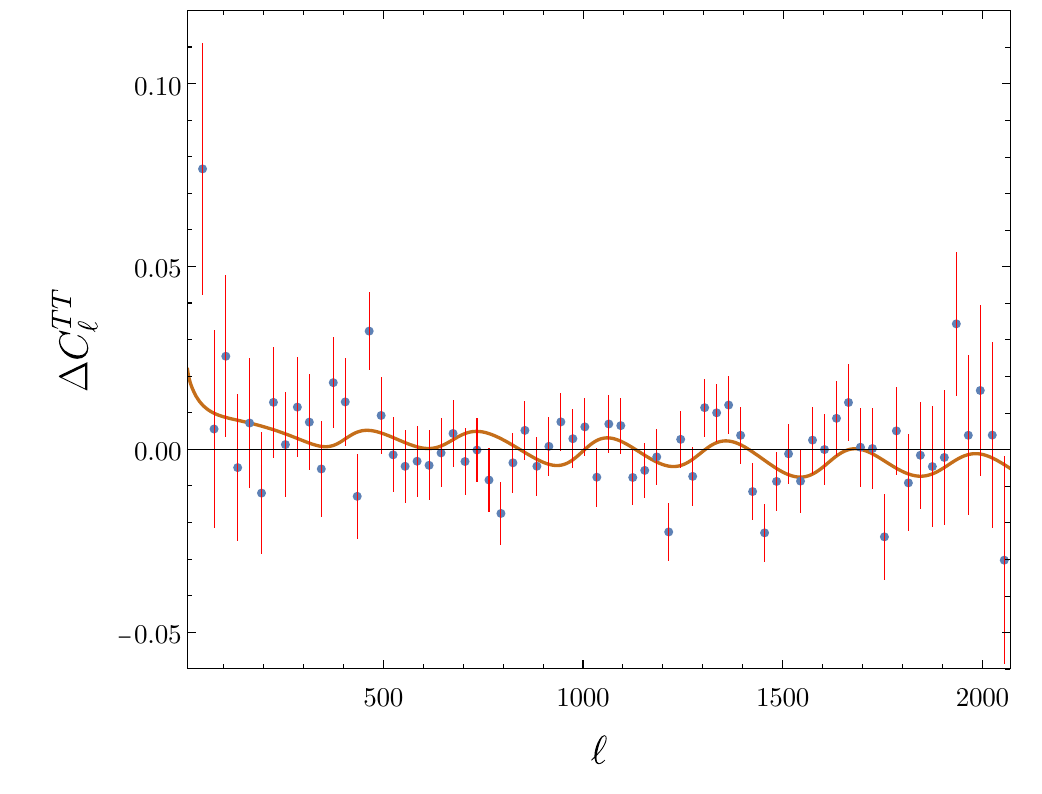}}\hfill
\subfigure[]{\includegraphics[scale=0.40]{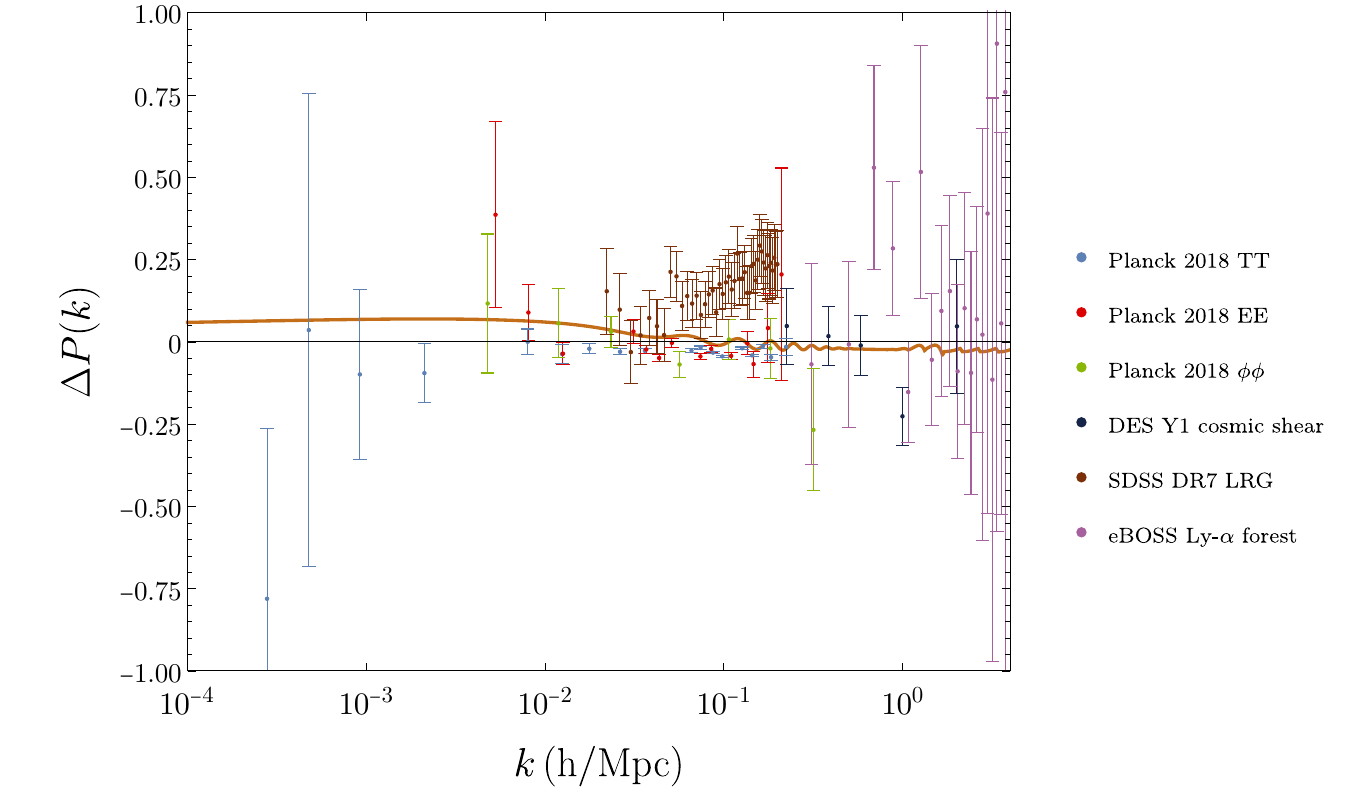}}
\caption{{\it Left:} Induced changes in the CMB temperature power spectrum in the ALP model using the best-fit values of the parameters in Table~\ref{tab:allbestfits} for the CMB alone. 
{\it Right:} Same for the linear matter power spectrum $P (k)$. In each case we define the fractional change $\Delta X\equiv \frac{X-X_{\Lambda\rm{CDM}}}{X_{\Lambda\rm{CDM}}}$, where $X$ is either $C^{TT}_{\ell}$ or $P(k)$. 
The CMB power spectrum data points are extracted from \url{http://pla.esac.esa.int/pla} and for the matter power spectrum from  \url{https://github.com/marius311/mpk_compilation}.
}
\label{fig:perturbations}
\end{figure}

\begin{figure}[ht!]
\centering
\subfigure[]{\includegraphics[scale=0.40]{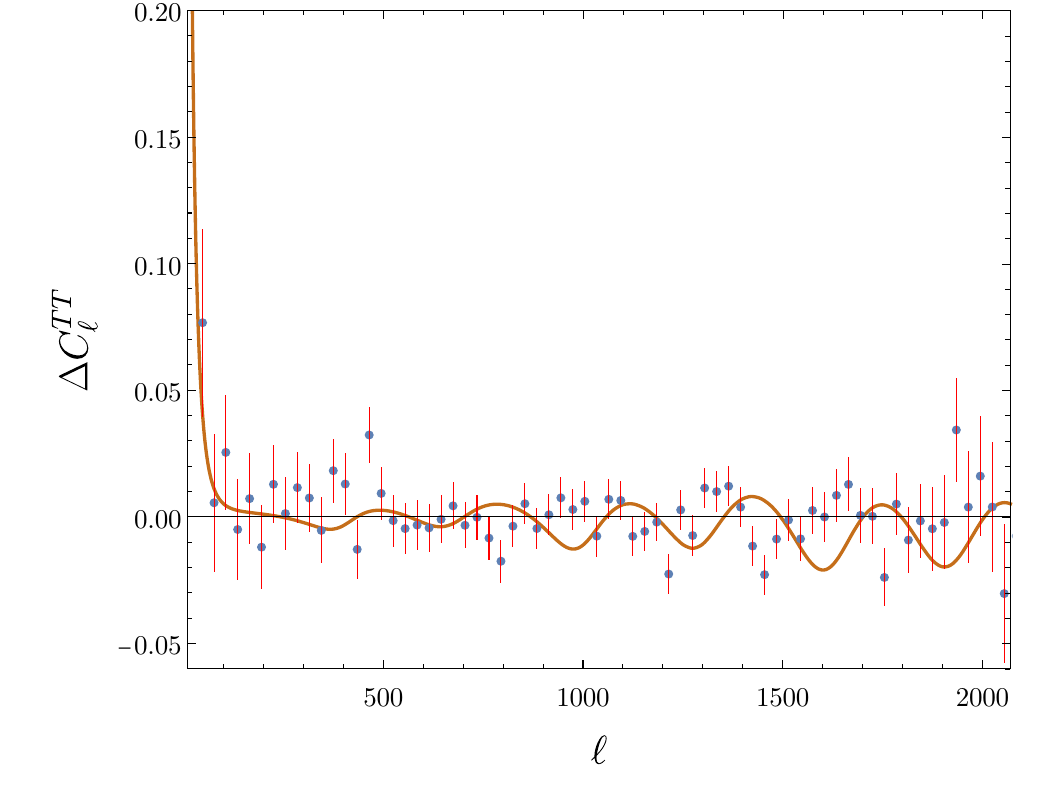}}\hfill
\subfigure[]{\includegraphics[scale=0.40]{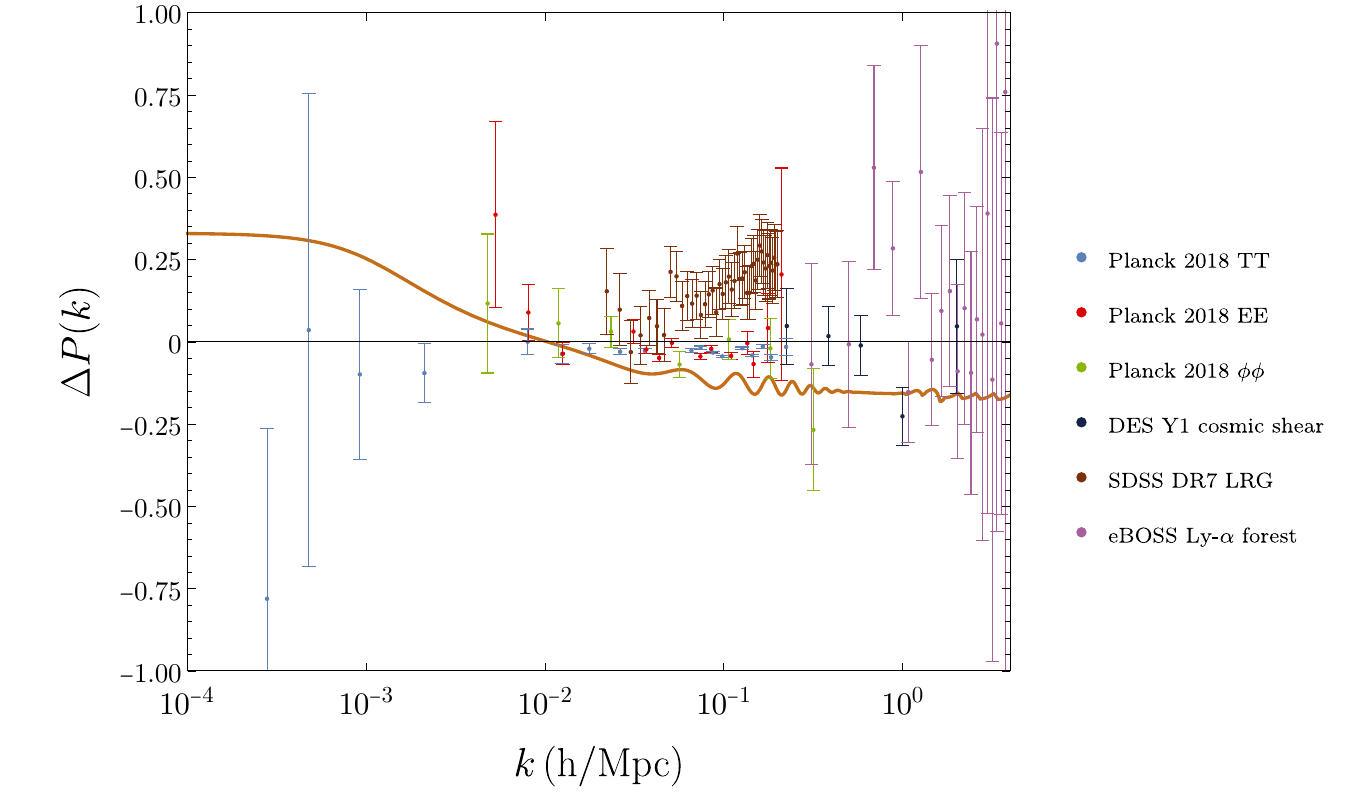}}
\caption{Same as in fig.~\ref{fig:perturbations} but for the benchmark point $a_{\rm osc} = 10^{-8}$, $a_{\rm DR}=0.003$, $\xi=0.7$. The rest of the parameters are fixed to their best-fit values in the $\Lambda$CDM model  for the CMB alone in ref.~\cite{Planck:2018vyg}.}
\label{fig:perturbations_benchmark}
\end{figure}

The numerical results obtained using Cobaya are summarized in Tables~\ref{tab:68_confidence_intervals}, \ref{tab:bestfith0s8}, and \ref{tab:allbestfits}. We use the Gelman-Rubin convergence criterion and run our Markov chains until we get a convergence of $R-1<0.07$ for each parameter. The posteriors distributions of the parameters $H_0$, $S_8$, $\log_{10}\xi$ and $\log_{10}a_{\rm DR}$ are shown in fig.~\ref{fig:posteriors}. In view of these results, it is clear that
 even though the model has the potential to resolve both the $S_8$ and $H_0$ tensions, it fails. According to Table~\ref{tab:68_confidence_intervals}, the values of $H_0$ and $S_8$ are compatible with those of $\Lambda$CDM. This is because Cobaya tends to minimize deviations of the ALP model from $\Lambda$CDM to reduce its additional contributions to the CMB temperature and matter power spectrum. This effect is illustrated comparing figs.~\ref{fig:perturbations} and \ref{fig:perturbations_benchmark}, where we show, for the best-fit values to the CMB alone in Table~\ref{tab:allbestfits} and for the reference point, the departure of the ALP model with respect to the data in the CMB temperature and matter power spectrum. Particularly, the matter power spectrum is reduced due to the 
 lack of DM at late times, after recombination. For comparison, the $\chi^2$ to the selected data points for the $\Lambda$CDM best-fit is $191.1$ while for the ALP best-fit and benchmark points is $192.7$ and $1101.2$, respectively. Nonetheless, it is worth noting that, according to Table~\ref{tab:bestfith0s8}, which shows the best-fit values of $H_0$ and $S_8$ and the corresponding $\chi^2$ for each dataset, in general the ALP model fits the CMB data slightly better than $\Lambda$CDM for the Planck and Planck+LSS datasets, as seen by comparing the respective $\chi^2_{\rm CMB}$.

\begin{figure}[ht!]
\includegraphics[width=10cm]{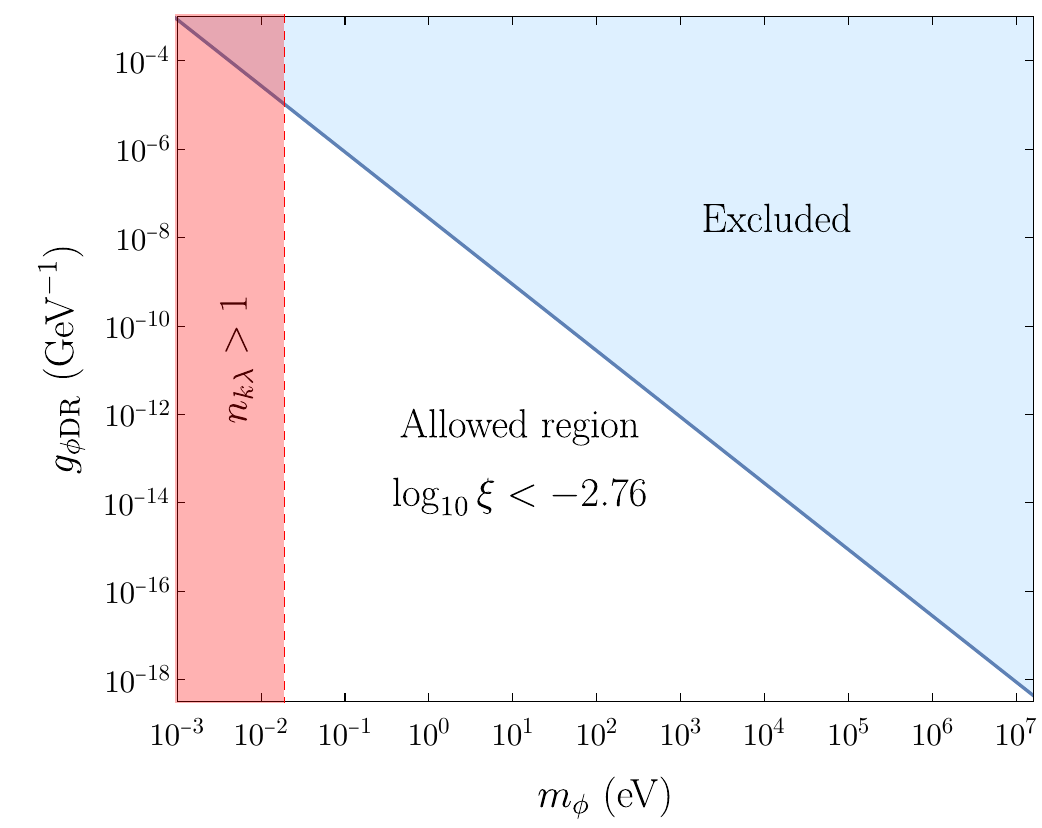}
\centering
\caption{Allowed region in the $m_{\phi}-g_{\phi\rm{DR}}$ plane after imposing  the 68\% confidence limits of Table~\ref{tab:68_confidence_intervals} and the approximation of small number of dark photons produced.} 
\label{fig:mass_vs_coupling}
\end{figure}

The other important result is that the fitted parameters in Table~\ref{tab:68_confidence_intervals} can provide a constraint on the ALP model parameters. In particular we obtain an upper bound $\log_{10}\xi<2.76$, which according to eq.~(\ref{ALPenergydensity}) can be translated into an upper bound for the ALP to DR coupling $g_{\phi\rm{DR}}$ as a function of the ALP mass
\eq{
g^2_{\phi\rm{DR}}=96\pi\xi\frac{H_0}{m^3_{\phi}}.
}
Substituting the values of $H_0$ and the upper bound of $\xi$ from the first column of Table~\ref{tab:68_confidence_intervals} we get
\eq{
\left(\frac{g_{\phi\rm{DR}}}{1~\rm{GeV^{-1}}}\right)^2<\left(\frac{m_{\phi}}{10^{-5}~\rm{eV}}\right)^{-3}.
}
However we still need to impose the lower bound on the ALP mass in eq.~(\ref{lowerboundALPmass}) that comes from imposing that  the number density of dark photons remains small. This expression depends on both $\xi$ and $a_{\rm DR}$. Unfortunately, Cobaya is not able to find the $68\%$ confidence limits for $a_{\rm DR}$. Nevertheless, in this model the ALP conversion to DR occurs during matter domination epoch so that $a_{\rm DR}>a_{\rm{eq}}\approx 10^{-3}$. This allows us to impose a sufficient condition
\eq{
\frac{m_{\phi}}{0.1~\rm eV}\geq \xi^{\frac{1}{4}}.
}
Finally, since we have an upper bound for $\xi$, it is enough to impose
\eq{
\frac{m_{\phi}}{0.1~\rm eV}\geq \left(10^{-2.76}\right)^{\frac{1}{4}}\rightarrow m_{\phi}>0.02~\rm eV.
}
As already announced, this lower bound is more stringent that the one that comes from fixing $a_{\rm osc}=10^{-8}$. The allowed region of the $m_{\phi}-g_{\phi\rm{DR}}$ plane compatible with all of our approximations is shown in fig.~\ref{fig:mass_vs_coupling}. Notice that cosmology excludes regions with large ALP masses and large couplings. For instance, for an ALP mass of order $m_\phi\sim 1$~eV, the coupling to DR should be less than $10^{-7}~\rm{GeV}^{-1}$ in order to be compatible with the data.

\begin{sidewaystable}

\small
\centering
\begin{tabular}{|c||c|c||c|c||c|c|c|c||c|c|c|c|}
\hline
\multicolumn{1}{|c||}{}&\multicolumn{4}{|c||}{}&\multicolumn{4}{c||}{$H_0$ [km/s/Mpc]}&\multicolumn{4}{c|}{$S_8$}\\\cline{2-13}
&\multicolumn{2}{c||}{$\log(\mathrm{Posterior)}$}&\multicolumn{2}{|c||}{$\chi^2_{\rm{CMB}}$}&\multicolumn{2}{|c|}{Mean}&\multicolumn{2}{|c||}{MAP}&\multicolumn{2}{|c|}{Mean}&\multicolumn{2}{|c|}{MAP}\\\cline{2-13}
&   $\Lambda$CDM  &   ALP &   $\Lambda$CDM  &   ALP &      $\Lambda$CDM  &   ALP &   $\Lambda$CDM  &   ALP &   $\Lambda$CDM  &   ALP &   $\Lambda$CDM  &   ALP      \\ \hline \hline

\textit{Planck} & 2804.9 & 2809.6 & 2765.6 & 2764.5 & 67.4$\pm$0.6 & 67.4$\pm$0.6 & 66.8 & 67.1 & 0.83$\pm$0.02 & 0.83$\pm$0.02 & 0.85 & 0.84    \\\hline
+LSS & 3875.5 & 3882.1 & 2774.8 & 2773.4 & 67.9$\pm$0.4 & 67.9$\pm$0.4 & 68.0 & 67.7 & 0.82$\pm$0.01 & 0.82$\pm$0.01 & 0.81 & 0.82   \\\hline
+SH0ES & 3891.9 & 3897.3 & 2777.0 & 2777.1 & 68.4$\pm$0.4 & 68.4$\pm$0.4 & 68.5 & 68.3 & 0.81$\pm$0.01 & 0.81$\pm$0.01 & 0.81 & 0.81    \\\hline

\end{tabular}

\caption{The $\log(\rm{Posterior})$ at the maximum a posteriori (MAP) point for the various datasets, along
with the $\chi^2$ of the MAP point of the Planck primary CMB alone $\chi^2_{\rm CMB}$ . We also show the mean value of $H_0$ from the MCMC chains, along with its value at the best-fit point; similarly for $S_8$. The error bars indicated on the mean are calculated from the covariance of the MCMC chains. To save space, we have used \textit{Planck} to refer to the \textit{Planck} primary CMB likelihood, and $+$LSS to denote $+\phi\phi$+BAO+SN+DES. }\label{tab:bestfith0s8}
\end{sidewaystable}

\begin{table} 
\centering
\begin{tabular}{ |l|  c c |c c| c c|}\hline
&\multicolumn{2}{c|}{\textit{Planck} primary CMB}&\multicolumn{2}{c|}{$+\phi\phi+$BAO+SN+DES}&\multicolumn{2}{c|}{+SH0ES}\\\cline{2-7}
 Parameter &  ALP&$\Lambda$CDM&  ALP&$\Lambda$CDM& ALP & $\Lambda$CDM\\\hline\hline

$\log(10^{10} A_\mathrm{s})$   &   3.05427   &   3.04904   &   3.04255   &   3.04192   &    3.05096   &   3.06036\\
$n_\mathrm{s}   $   & 0.96178   &   0.96148  &   0.96558  &   0.96856   &  0.97033  &  0.97109\\
$100\Omega_\mathrm{b} h^2$   &  2.23475    &   2.23161   &   2.24125   &  2.24453   &   2.25612   &   2.25862\\
$\Omega_\mathrm{c} h^2$   &  0.12046 &   0.12146   &   0.11919   &   0.11853  &   0.11793  &   0.11757\\
$\tau_\mathrm{reio}$   &  0.05712    &    0.05241  &   0.06020  &   0.05466  &   0.06020   &   0.06505\\
$100\theta_\mathrm{s}$   &  1.04180   &   1.04175   &   1.04196  &   1.04191   &   1.04202  &    1.04213\\
$\log_{10}\xi$   &  -3.21188   & ---     &   -4.92677  & ---     &  -2.54401   & ---  \\
$\log_{10}(a_{\rm DR}) $   &  -2.96924   & ---     &   -1.03993   & ---     &   -0.04431  & --- 
 \\\hline\hline
$H_0                       $ & 67.14     & 66.77     & 67.72  & 67.97   & 68.31  & 68.52 \\

$\Omega_\mathrm{m}         $ & 0.318 & 0.324 & 0.310  & 0.307& 0.302 & 0.300\\

$S_8                       $& 0.840& 0.849& 0.821 & 0.815 & 0.811& 0.810\\
\hline

\end{tabular}
\caption{The values of the base $\Lambda$CDM and ALP model parameters, along with the derived parameters $H_0$, $\Omega_{\rm m}$ and $S_8$ at the best-fit points for each analysis.}\label{tab:allbestfits}
\end{table}

\section{Conclusions}\label{conclusions}
In this work we have considered an ALP particle with a cosine potential that also couples to a DR sector. The global symmetry responsible of the existence of the ALP is spontaneously broken before inflation (the pre-inflationary scenario), meaning that in radiation domination era the ALP is an homogeneous field that behaves as a classical condensate. Assuming a standard cosmological histoy, the coherent oscillations of the ALP around one of the minima of its potential generate DM. The onset of these oscillations is model dependent. However, once the ALP starts oscillating, it can be described in a model independent fashion.

Due to the coupling of the ALP to DR, the coherent oscillations of the ALP not only produce DM, but also dark radiation. We have calculated explicitly the energy density of DR using the WKB approach and the stationary phase approximation assuming that the ALP can be considered as a background field and including later the backreaction on the ALP by using energy conservation. It turns out that the later the ALP oscillations start the most DR is produced, being proportional to the ratio $a_{\rm osc}/H_{\rm osc}$.

We have implemented this model in the software CLASS and perform a Bayesian analysis using Cobaya to explore its cosmological implications. Using CMB data and the latest DES-Y3 $S_8$ and SH0ES $H_0$ constraints we tried to elucidate whether this model can indeed solve the $H_0$ and $S_8$ tensions. However, according to the bayesian analysis performed by Cobaya, our model is not able to significantly deviate from $\Lambda$CDM to relax both tensions. Specifically, this is because the deviations of the ALP with respect to the matter power spectrum data $P(k)$ tend to be large due to the lack of DM at late times. This can be seen comparing figs.~\ref{fig:perturbations} and \ref{fig:perturbations_benchmark} where we show the deviations of the ALP model with respect to the data in the CMB temperature and matter power spectrum for the best-fit values to the CMB alone in Table~\ref{tab:allbestfits} and for the benchmark point. However, according to the parameter $\chi^2_{\rm CMB}$ in Table~\ref{tab:bestfith0s8}, in general terms our model fits the CMB data slightly better than $\Lambda$CDM  when considering the Planck primary CMB likelihood and LSS datasets.

Another significant outcome of this work, should the cosmological tensions ultimately resolve, is the constraint cosmology imposes on the coupling of ALPs to dark radiation if ALPs constitute the sole component of Dark Matter. The parameter space region consistent with the approximation of a small number of produced dark photons is illustrated in fig.~\ref{fig:mass_vs_coupling}. Note that the data provide an upper bound on the ALP coupling to DR as a function of its mass. For example, for a mass of order $1~\rm{eV}$, the coupling can be as large as $g_{\phi\rm{DR}}\approx 10^{-7}~\rm{GeV}^{-1}$. Hence, taking $f_{\phi}\sim g_{\phi\rm{DR}}^{-1}$, we get $f_{\phi}\sim 10^{7}~\rm{GeV}$.

In our analysis, we have focused exclusively on ALP couplings to new gauge bosons; however, future studies could extend this framework by considering potential couplings to fermions as well.

\section*{Acknowledgments}
The research of JMPP and VS is supported by the Generalitat
Valenciana PROMETEO/2021/083. The research of VS is also supported by the Proyecto Consolidacion CNS2022-135688, and the Ministerio de Ciencia e
Innovacion PID2020-113644GB-I00 and the {\it Severo Ochoa} project CEX2023-001292-S funded by MCIU/AEI. 
\printbibliography  

\end{document}